\documentclass[altaffilletter,aps,nofootinbib,twocolumn,prd,eqsecnum,preprintnumbers,superscriptaddress,10pt,floatfix]{revtex4-2}
\pdfoutput=1
\usepackage{graphicx}
\usepackage{comment}
\usepackage{amsmath}
\usepackage{subfigure}
\usepackage{amssymb}
\usepackage{amsfonts}
\usepackage{mathtools}
\usepackage{amssymb}
\usepackage{enumerate}
\usepackage{xcolor}
\usepackage{bm}
\usepackage{mathrsfs}
\usepackage{epstopdf}
\usepackage{url}
\usepackage{footnote}
\usepackage{textcomp}
\usepackage{dsfont}
\usepackage{ulem}
\usepackage{hyperref}
\usepackage{enumerate}   
\usepackage{appendix}
\usepackage{textcomp}
\usepackage{tipa}

\makeatletter
\newcommand*{\rom}[1]{\expandafter\@slowromancap\romannumeral #1@}
\makeatother

\begin{document}

\title{Quasinormal-mode redshifts of black holes in galactic halos with radial pressures}

\author{Che-Yu Chen}
\email{b97202056@gmail.com}
\affiliation{RIKEN iTHEMS, Wako, Saitama 351-0198, Japan}

\author{Hassan Hassanabadi
}
\email{hassanhassanabadi@mail.fresnostate.edu}
\affiliation{ Physics Department, California State University, Fresno, CA 93740, USA}

\author{Soroush Zare%$^{b,d,h,k}$
}
\email{soroush.z.zare@helsinki.fi}
\affiliation{ Helsinki Institute of Physics, University of Helsinki, P.O. Box 64, FI-00014 Helsinki, Finland}

\begin{abstract}

We investigate the black-hole quasinormal mode redshifts induced by a galactic dark-matter halo with nonvanishing radial pressure. The halo is described by an anisotropic fluid with a generalized density profile and a constant radial equation-of-state parameter \(w\). We construct the corresponding static, spherically symmetric spacetime and study the resulting corrections to the unstable light ring and quasinormal-mode spectrum. For dilute halos and \(w\neq -1\), the light-ring angular frequency and Lyapunov exponent acquire the same leading-order gravitational redshift, implying a universal shift of the real and imaginary parts of the quasinormal frequencies. The magnitude of this effect depends on both the halo structure and the radial pressure. This universal redshift relation holds not only in the case of $w=0$ that has been discovered in the literature but also in a much wider scenario, supporting its importance in black hole spectroscopy. On the other hand, the case \(w=-1\) is qualitatively different: the factor that corresponds to the leading common redshift vanishes, the corrections are further suppressed by the hierarchy between the black-hole and halo scales, and the universal relation is broken, with the dominant contribution controlled by the inner halo profile. Our numerical computations of scalar-field quasinormal mode spectra confirm the analytic eikonal results.
\end{abstract}

\maketitle

\section{Introduction}\label{sec.intro}

Black holes (BHs) have evolved from theoretical solutions of the Einstein equations into precision laboratories for strong-field gravity. Gravitational-wave (GW) detections of compact-binary coalescences, stellar-orbit measurements in the Galactic center, and horizon-scale imaging of M87* and Sgr A* now provide complementary observational evidence for astrophysical BHs and access to their dynamics over widely separated scales \cite{LIGOScientific:2016aoc,LIGOScientific:2020ibl,KAGRA:2021vkt,LIGO:2021ppb,Ghez:1998ph,genzel2010galactic,EventHorizonTelescope:2019dse,EventHorizonTelescope:2022wkp}. Of particular relevance to tests of gravity is the post-merger ringdown, during which the remnant relaxes through a superposition of damped oscillations characterized by quasinormal modes (QNMs) \cite{Kokkotas:1999bd,Berti:2009kk,Konoplya:2011qq,Vishveshwara:1970zz}. In vacuum general relativity, the QNM spectrum of a stationary BH is fixed by its intrinsic parameters and therefore underlies the program of BH spectroscopy, providing a probe of the remnant geometry and a framework for consistency tests of general relativity  \cite{Berti:2005ys,Giesler:2019uxc,Baibhav:2023clw,destounis2024black,Cardoso:2019rvt,Berti:2025hly,Franchini:2023eda,Vishveshwara:1970cc,Press:1971wr,chandrasekhar1983mathematical}. Future space-based observations of massive BH systems are expected to extend such tests to the mHz band, making small systematic shifts of the ringdown spectrum increasingly relevant \cite{amaro2017laser,Baibhav:2020tma,Piro:2022zos,LISA:2022yao,Deng:2025qhx}.

Astrophysical BHs, however, are not isolated vacuum objects. There is compelling gravitational evidence for dark matter (DM), although its microscopic nature remains unknown and searches for nongravitational interactions have so far yielded no confirmed detection \cite{Freese:2008cz,Navarro:1995iw,Clowe:2006eq,Bertone:2004pz,Kahlhoefer:2017dnp,PerezdelosHeros:2020qyt}. Supermassive BHs reside in galactic nuclei and are therefore embedded in extended matter distributions; stellar dynamics around Sgr A*, megamaser observations, and empirical correlations between central BHs and their host galaxies all emphasize the intimate connection between the central compact object and the surrounding galactic structure \cite{Ghez:2008ms,Gillessen:2008qv,Miyoshi:1995da,Kormendy:2013dxa,Ferrarese:2000se,Gebhardt:2000fk}. Galaxies themselves are embedded in DM halos, while the distribution near a central BH can be substantially reshaped by BH growth and spike formation, capture of low-angular-momentum particles, stellar scattering, mergers, and long-term galactic evolution \cite{Navarro:1996gj,Salucci:2018hqu,Gondolo:1999ef,Sadeghian:2013laa,Merritt:2002vj,Merritt:2003qk,Ullio:2001fb,Bertone:2005hw}. Such matter can influence compact-binary dynamics as well as the generation and propagation of GWs, so a first-principle treatment requires the spacetime geometry sourced by the environment rather than a purely Newtonian correction superposed on a vacuum BH metric \cite{Barack:2018yly,Eda:2013gg,Macedo:2013qea,Barausse:2014tra,Baibhav:2019rsa,Seoane:2021kkk,Cardoso:2019rou,Kavanagh:2020cfn,Tamanini:2019usx,Chowdhury:2025tpt,Dosopoulou:2025jth,Hassanabadi:2026wgc}.

A central ingredient of this program is the choice of the galactic density profile. Hernquist-type distributions provide useful models for elliptical galaxies and galactic bulges, whereas the Navarro-Frenk-White (NFW) profile is motivated by cosmological structure-formation simulations; Jaffe and King profiles furnish further widely used alternatives \cite{Navarro:1995iw,hernquist1990analytical,Jaffe:1983iv,King:1962wi}. These distributions can be placed within a generalized $\{\alpha,\beta,\gamma\}$ family, in which $\gamma$ determines the inner logarithmic slope, $\beta$ controls the outer falloff, and $\alpha$ specifies the sharpness of the transition between the two regimes \cite{Graham:2005xx,Taylor:2002zd,dekel2017dark,Zhao:1995cp}. The presence of a BH also alters the innermost halo, motivating prescriptions that suppress the density close to the horizon or to the capture region rather than extrapolating a galactic profile unchanged into the strong-field domain \cite{Gondolo:1999ef,Sadeghian:2013laa,Kavanagh:2020cfn,Hassanabadi:2026wgc,Cardoso:2021wlq,Maeda:2024tsg,Nampalliwar:2021tyz,Capozziello:2023tbo}. A fully relativistic realization was developed by Cardoso et al. by extending the Einstein-cluster construction to include a central BH; the matter is represented by an anisotropic effective fluid with nonzero tangential pressure and vanishing radial pressure, $p_r=0$ \cite{Cardoso:2021wlq,Einstein:1939ms,Geralico:2012jt}. This framework subsequently served as a basis for numerical BH solutions with Hernquist, NFW, and more general halo profiles \cite{Konoplya:2022hbl,Figueiredo:2023gas,Speeney:2024mas,Pezzella:2024tkf,Bhowmik:2026owi}.

The influence of these environments on BH perturbations has revealed a particularly simple leading-order structure. For a BH embedded in a sufficiently dilute Hernquist-type environment, Cardoso et al. \cite{Cardoso:2021wlq} found that the dominant modification of the light-ring dynamics and of the complex QNM frequencies can be interpreted as a gravitational redshift generated by the surrounding galactic potential; subleading effects remain in GW propagation and become more relevant away from the high-frequency regime \cite{Cardoso:2008bp,Berti:2009kk,Jansen:2017oag,Cardoso:2017soq,Berti:2005ys,amaro2017laser,Seoane:2021kkk,ET:2019dnz}. 
Pezzella et al. extended the QNM calculation to numerically constructed Hernquist- and NFW-type BHs and showed that, for dilute halos, axial modes exhibit an approximately universal redshift governed by the central Newtonian potential, with both the oscillation frequency and damping rate shifted at leading order by the halo compactness \cite{Pezzella:2024tkf}. 
More recently, Bhowmik et al. generalized this picture to a broad class of $\{\alpha,\beta,\gamma\}$ halos and showed that the redshift also retains information about the radial distribution of matter: the inner slope can dominate for centrally concentrated profiles, while $\alpha$ and $\beta$ introduce additional profile-dependent corrections \cite{Bhowmik:2026owi}. Thus, the emerging picture is that compactness controls the overall scale of the environmental correction, while the internal structure of the halo can determine how that correction is imprinted on the ringdown.

The relativistic BH-halo constructions underlying this sequence of QNM-redshift studies share, however, an important structural restriction: the Einstein-cluster description sets the radial pressure to zero, $p_r=0$, leaving only the tangential stresses of the effective anisotropic fluid \cite{Cardoso:2021wlq}. This motivates asking how robust the environmental-redshift picture remains when the halo carries an independent radial stress. Instead of being a purely phenomenological consideration, DM halos with non-zero radial pressure can be motivated by various DM models, including the collisional features of fluids in Self-Interacting Dark Matter models \cite{Tulin:2017ara} and the quantum pressure in Fuzzy Dark Matter models \cite{Eberhardt:2025caq}. In the present work, we address this question for a static and spherically symmetric BH embedded in a generalized $\{\alpha,\beta,\gamma\}$ halo by introducing a radial equation of state $p_r=w\rho$; unlike the mass function, the redshift function then acquires an explicit dependence on $w$ through the radial Einstein equation, discussed in Sec.~\ref{sec.spacetime}. We determine the resulting light-ring quantities and QNM redshifts and show that, for $w\neq-1$, the real and imaginary parts of the eikonal QNM frequencies acquire the same leading environmental redshift, now controlled by both the halo profile and the radial-pressure parameter, whereas $w=-1$ constitutes a distinct case in which the leading contribution is further suppressed by the hierarchy between the BH and halo scales, discussed in Sec.~\ref{sec:phqnm}. The $w=0$ limit recovers the established zero-radial-pressure construction, thereby isolating the role of radial stresses in the environmental modification of BH ringdown.

The paper is organized as follows. In Sec.~\ref{sec.spacetime}, we introduce the setup of the BH-DM system considered in this work. With the radial pressure modeled by $p_r=w\rho$, we solve the mass function (Sec.~\ref{subsec.mass}) and the redshift function (Sec.~\ref{subsec.redshift}) using the Einstein equations. In Sec.~\ref{subsec.root}, we show that as long as the halo is sufficiently dilute, no extra event horizon would appear. In Sec.~\ref{sec:phqnm}, we discuss the modifications of the light ring structure (radius, angular frequency, the Lyapunov exponent) and the QNM redshifts induced by the surrounding halo. Finally, we conclude in Sec.~\ref{sec:conclusion}.

\section{Spacetime setup}\label{sec.spacetime}

We consider a static and spherically symmetric spacetime whose line element reads
\begin{equation}
ds^2=-A(r)dt^2+\frac{dr^2}{1-2M(r)/r}+r^2\left(d\theta^2+\sin^2\theta d\varphi^2\right)\,,
\end{equation}
where $M(r)$ and $A(r)$ are the mass function and the redshift function, respectively. 

Suppose that the energy-momentum tensor $T_{\mu\nu}$ of the fluid that supports the halo takes the form
\begin{align}
\rho&\equiv -T^t_t\,,\quad p_r\equiv T^r_r\,,\nonumber\\
p_t&\equiv T^\theta_\theta=T^\varphi_\varphi\,.
\end{align}
The $tt$ and the $rr$ components of the Einstein equations read
\begin{align}
\kappa\rho&=\frac{2M'(r)}{r^2}\,,\label{eomM}\\
\kappa p_r&=\frac{1}{r^3}\left[\frac{A'(r)}{A(r)}\left(r^2-2rM(r)\right)-2M(r)\right]\,,\label{eompr}
\end{align}
where $\kappa\equiv 8\pi G$ and the prime denotes the derivative with respect to $r$. From now on, we set $G=1$. 

In this work, we will focus on how the presence of the radial pressure $p_r$ could leave imprints on the QNM redshifts. To this end, we will parameterize the radial pressure $p_r$ as
\begin{equation}
p_r=w\rho\,,
\end{equation}
where $w$ is a constant equation of state. Combining with Eqs.~\eqref{eomM} and \eqref{eompr}, one can obtain the following equation
\begin{align}
\frac{A'(r)}{A(r)}=\frac{2wr M'(r)+2M(r)}{r^2-2rM(r)}\,.\label{eomA}
\end{align}
This equation explicitly depends on the equation of state $w$ on its right-hand side. Note that when $w=-1$, one gets the special case in which $A(r)=c_0\left(1-2M(r)/r\right)$ with $c_0$ a constant. The constant $c_0$ can be set to one by a constant rescaling of the time variable $t$.

We consider the following DM energy density
\begin{equation}\label{DMDensityProfile}
	\rho(r) = \frac{(3-\gamma){\rm M}_{\rm DM}^{\rm tot}}{4\pi r_{\rm s}^{3}}\left(\frac{r}{r_{\rm s}}\right)^{-\gamma}\left[1+\left(\frac{r}{r_{\rm s}}\right)^{\alpha}\right]^{\frac{\gamma-\beta}{\alpha}}b(r)\,,
\end{equation}
where $r_{\rm s}$ denotes the scale radius, ${\rm M}_{\rm DM}^{\rm tot}$ is the total DM mass of the halo, and $\alpha>0$, $\beta$, and $\gamma$ are dimensionless parameters that characterize the specific density profile. Note that we consider $\beta>3$ for a finite DM mass \cite{Konoplya:2022hbl,Hassanabadi:2026wgc}, and $0\le\gamma<2$. Compared with the original model, the extra factor $b(r)$ is introduced such that the density distribution vanishes at the horizon $r_0=2M_\textrm{BH}$ while recovering the usual distribution in the far zone. A convenient choice for $b(r)$ is
\begin{equation}
b(r)=1-\frac{r_0}{r}\,.
\end{equation}
Given the energy density \eqref{DMDensityProfile}, the metric functions $A(r)$ and $M(r)$ can be determined by solving Eqs.~\eqref{eomM} and \eqref{eomA}.

\subsection{The mass function $M(r)$}\label{subsec.mass}

We first discuss the mass function $M(r)$, which can be obtained by solving Eq.~\eqref{eomM}, or equivalently, 
\begin{equation}
M'(r)=4\pi r^2\rho(r)\,,
\end{equation}
with the condition $M(r_0)=r_0/2$. An analytic and exact solution for $M(r)$ can be obtained:
\begin{widetext}
\begin{align}
M(r)=\frac{r_0}{2}+&{\rm M}_{\rm DM}^{\rm tot}y_0^{3-\gamma}\left[\left(\frac{\gamma-3}{\gamma-2}\right){}_2F_1\left(\frac{2-\gamma}{\alpha},\frac{\beta-\gamma}{\alpha};\frac{2+\alpha-\gamma}{\alpha};-y_0^\alpha\right)-{}_2F_1\left(\frac{3-\gamma}{\alpha},\frac{\beta-\gamma}{\alpha};\frac{3+\alpha-\gamma}{\alpha};-y_0^\alpha\right)\right]\nonumber\\
-&{\rm M}_{\rm DM}^{\rm tot}y^{3-\gamma}\left[\frac{y_0}{y}\left(\frac{\gamma-3}{\gamma-2}\right){}_2F_1\left(\frac{2-\gamma}{\alpha},\frac{\beta-\gamma}{\alpha};\frac{2+\alpha-\gamma}{\alpha};-y^\alpha\right)-{}_2F_1\left(\frac{3-\gamma}{\alpha},\frac{\beta-\gamma}{\alpha};\frac{3+\alpha-\gamma}{\alpha};-y^\alpha\right)\right]\,,\label{exactmass}
\end{align}
\end{widetext}
where we have defined $y\equiv r/r_{\rm s}$ and $y_0\equiv r_0/r_{\rm s}$. It turns out that $\gamma\ne2$ otherwise the mass is ill-defined. In addition, it should be emphasized that the mass function \eqref{exactmass} is completely exact and it does not depend on the equation of state $w$.

In the large-$r$ limit where $y\gg1$, the mass function can be approximated as
\begin{widetext}
\begin{align}
M(r)\approx&\,\frac{r_0}{2}+{\rm M}_{\rm DM}^{\rm tot}\frac{\Gamma\left(\frac{\beta-3}{\alpha}\right)\Gamma\left(\frac{3+\alpha-\gamma}{\alpha}\right)}{\Gamma\left(\frac{\beta-\gamma}{\alpha}\right)}\nonumber\\&-{\rm M}_{\rm DM}^{\rm tot}y_0\left(\frac{\gamma-3}{\gamma-2}\right)\frac{\Gamma\left(\frac{\beta-2}{\alpha}\right)\Gamma\left(\frac{2+\alpha-\gamma}{\alpha}\right)}{\Gamma\left(\frac{\beta-\gamma}{\alpha}\right)}-\frac{{\rm M}_{\rm DM}^{\rm tot}}{\gamma-2}y_0^{3-\gamma}\left[1+\mathcal{O}(y_0^\alpha)\right]\,.\label{massbigr}
\end{align}
\end{widetext}
In the usual galactic configuration, the DM scale is much larger than the horizon radius, i.e., $r_{\rm s}\gg r_0$ or $y_0\ll1$. In such a scenario, the two terms in the second line of Eq.~\eqref{massbigr} are negligible. Therefore, the total mass read at asymptotic region is approximated by
\begin{equation}
M(r)\approx\frac{r_0}{2}+{\rm M}_{\rm DM}^{\rm tot}\frac{\Gamma\left(\frac{\beta-3}{\alpha}\right)\Gamma\left(\frac{3+\alpha-\gamma}{\alpha}\right)}{\Gamma\left(\frac{\beta-\gamma}{\alpha}\right)}\,.
\end{equation}
In particular, when $\beta=\alpha+3$, the coefficient associated with ${\rm M}_{\rm DM}^{\rm tot}$ in the second term is one, and the total mass is given by $M(r)\approx r_0/2+{\rm M}_{\rm DM}^{\rm tot}$.

Following the procedure in Ref.~\cite{Konoplya:2022hbl}, we rewrite the mass function in terms of a new function $\mathcal{B}(y)$ as follows
\begin{equation}
1-\frac{2M(r)}{r}=\left(1-\frac{r_0}{r}\right)\mathcal{B}(y)\,.\label{mbpara}
\end{equation}
Then, one can further rewrite $\mathcal{B}(y)$ as \cite{Konoplya:2022hbl}
\begin{equation}
\mathcal{B}(y)=1-\frac{2{\rm M}_{\rm DM}^{\rm tot}}{r_{\rm s}}\widetilde{\mathcal{B}}(y)\,,\label{bybtilde}
\end{equation}
where $\widetilde{\mathcal{B}}(y)$, in the limit $y_0\ll1$, can be expressed as \cite{Konoplya:2022hbl}
\begin{equation}
\widetilde{\mathcal{B}}(y)=y^{2-\gamma}{}_2F_1\left(\frac{3-\gamma}{\alpha},\frac{\beta-\gamma}{\alpha};\frac{3+\alpha-\gamma}{\alpha};-y^\alpha\right)\,.\label{Bwidetildegen}
\end{equation}
Note that if $\gamma<2$, this function satisfies $\widetilde{\mathcal{B}}(0)=0$.

\subsection{The redshift function $A(r)$}\label{subsec.redshift}

To proceed, we rewrite the redshift function $A(r)$ as
\begin{equation}
A(r)=\left(1-\frac{r_0}{r}\right)\mathcal{A}(y)
\end{equation}
to ensure that the redshift function also vanishes at the horizon $r_0$. Then, by taking $y_0\ll1$, Eq.~\eqref{eomA} can be rewritten as
\begin{equation}
\frac{\mathcal{A}_{,y}}{\mathcal{A}}+\frac{wy\mathcal{B}_{,y}+(1+w)\left(\mathcal{B}-1\right)}{y\mathcal{B}}=0\,.\label{aab}
\end{equation}
Similar to the procedure in Ref.~\cite{Konoplya:2022hbl}, we approximate 
\begin{equation}
\mathcal{A}(y)\approx 1-\frac{2{\rm M}_{\rm DM}^{\rm tot}}{r_{\rm s}}\widetilde{\mathcal{A}}(y)\,,
\end{equation}
such that Eq.~\eqref{aab} can be recast as
\begin{equation}
\widetilde{\mathcal{A}}_{,y}=\frac{-\widetilde{\mathcal{B}}(w+1)-wy\widetilde{\mathcal{B}}_{,y}}{y}\,.\label{aab2}
\end{equation}
This equation has an analytic solution given by
\begin{equation}
\widetilde{\mathcal{A}}(y)=-(w+1)H(y)-w\widetilde{\mathcal{B}}(y)\,,\label{Awidetildeana}
\end{equation}
where
\begin{widetext}
\begin{align}
H(y)=&y^{2-\gamma}\left[\frac{\Gamma\left(\frac{2-\gamma}{\alpha}\right)\Gamma\left(\frac{3+\alpha-\gamma}{\alpha}\right)}{\Gamma\left(\frac{3-\gamma}{\alpha}\right)\Gamma\left(\frac{2+\alpha-\gamma}{\alpha}\right)}{}_2F_1\left(\frac{2-\gamma}{\alpha},\frac{\beta-\gamma}{\alpha};\frac{2+\alpha-\gamma}{\alpha};-y^\alpha\right)-{}_2F_1\left(\frac{3-\gamma}{\alpha},\frac{\beta-\gamma}{\alpha};\frac{3+\alpha-\gamma}{\alpha};-y^\alpha\right)\right]\nonumber\\
+&\frac{(\gamma-3)\Gamma\left(\frac{\beta-2}{\alpha}\right)\Gamma\left(\frac{2-\gamma}{\alpha}\right)}{\alpha\Gamma\left(\frac{\beta-\gamma}{\alpha}\right)}\,.\label{H}
\end{align}
\end{widetext}
Note that the constant term in the second line is added in order to ensure that the large-$r$ expression of the redshift function is the same as that of the $g^{rr}$ component, i.e., 
\begin{equation}
A(r)\approx 1-\frac{r_0+2{\rm M}_{\rm DM}^{\rm tot}\frac{\Gamma\left(\frac{\beta-3}{\alpha}\right)\Gamma\left(\frac{3+\alpha-\gamma}{\alpha}\right)}{\Gamma\left(\frac{\beta-\gamma}{\alpha}\right)}}{r}\,.
\end{equation}
In addition, it should be emphasized that although the functions $\widetilde{\mathcal{A}}(y)$ and $H(y)$ given in Eqs.~\eqref{Awidetildeana} and \eqref{H} are exact solutions to Eq.~\eqref{aab2}, the resulting redshift function $A(r)$ is not exact to the entire system. This is because in the derivation of Eq.~\eqref{aab2}, the assumption $y_0\ll1$ has been applied. 

Using the function $H(y)$ given in Eq.~\eqref{H} and recalling that $\widetilde{\mathcal{B}}(0)=0$ when $\gamma<2$, one can obtain $\widetilde{\mathcal{A}}(0)=-(w+1)H(0)$ where
\begin{equation}
H(0)=\frac{(\gamma-3)\Gamma\left(\frac{\beta-2}{\alpha}\right)\Gamma\left(\frac{2-\gamma}{\alpha}\right)}{\alpha\Gamma\left(\frac{\beta-\gamma}{\alpha}\right)}\,,\label{At0general}
\end{equation}
which is generically not zero. Therefore, $w=-1$ represents a special case where $\widetilde{\mathcal{A}}(0)=0$. 

Before closing this section, we would like to mention that in the Hernquist-type model with $(\alpha,\beta,\gamma)=(1,4,1)$, the redshift function constructed from Eqs.~\eqref{Awidetildeana} and \eqref{H} is
\begin{equation}
A(r)\approx\left(1-\frac{r_0}{r}\right)\left(1-\frac{2{\rm M}_{\rm DM}^{\rm tot}}{r+r_{\rm s}}-\frac{2w{\rm M}_{\rm DM}^{\rm tot}r_{\rm s}}{\left(r+r_{\rm s}\right)^2}\right)\,,
\end{equation}
which is valid when $r_0\ll r_{\rm s}$. When $w=0$, one gets
\begin{equation}
A(r)\approx\left(1-\frac{r_0}{r}\right)\left(1-\frac{2{\rm M}_{\rm DM}^{\rm tot}}{r+r_{\rm s}}\right)\,,
\end{equation} 
reproducing exactly the result in Ref.~\cite{Konoplya:2022hbl}.

\subsection{Absence of additional roots of the radial metric function}\label{subsec.root}

In this subsection, we would like to discuss the possibility of having extra roots of the radial metric function, which is related to the possible existence of extra horizons, an unwanted feature in the BH-halo setup. It turns out that as long as the halo is sufficiently dilute, the entire system can only have one horizon at $r=r_0$.

For $\alpha>0$, $\beta>3$, $0\leq\gamma<2$, and $M(r_0)=r_0/2$, define
$h(x)\equiv x^{2-\gamma}(1+x^\alpha)^{-(\beta-\gamma)/\alpha}$.  The exact
mass function \eqref{exactmass} can then be written compactly as
\begin{equation}
 \begin{aligned}
 M(r)=\frac{r_0}{2}+(3-\gamma){\rm M}_{\rm DM}^{\rm tot}
 \int_{y_0}^{y}h(x)\left(1-\frac{y_0}{x}\right)dx\,,
 \end{aligned}
 \label{eq:mass}
\end{equation}
where  $x$ is the integration variable. Then,  the exact form of function $\widetilde{\mathcal{B}}(y)$ in Eq.~\eqref{bybtilde} can be written as
\begin{equation}
 \widetilde{\mathcal{B}}(y)=\frac{3-\gamma}{y-y_0}
 \int_{y_0}^{y}h(x)\left(1-\frac{y_0}{x}\right)dx \,.
 \label{eq:F}
\end{equation}

At the prescribed horizon $r=r_0$, the numerator and denominator of $\widetilde{\mathcal{B}}$ vanish. Using L'Hopital's rule, one gets $\widetilde{\mathcal{B}}(y_0)=0$ because the integrand contains
$1-y_0/x$, and consequently $\mathcal{B}(y_0)=1$. Thus, $r=r_0$ is not a zero of $\mathcal{B}(y)$.

For $0<y<y_0$, the integrand in Eq.~\eqref{eq:F} is negative, while reversing
the integration limits makes its integral positive; since $y-y_0<0$, one has
$\widetilde{\mathcal{B}}(y)<0$. Hence
\begin{equation}
\mathcal{B}(y)=1-\frac{2{\rm M}_{\rm DM}^{\rm tot}}{r_{\rm s}}\widetilde{\mathcal{B}}(y)>1\,,\qquad 0<r<r_0\,,
 \label{eq:inside}
\end{equation}
so there is no extra root of $\mathcal{B}(y)$ inside the horizon $r_0$. This statement is exact
and does not require either $y_0\ll1$ or ${\rm M}_{\rm DM}^{\rm tot}/r_{\rm s}\ll1$.

For $y>y_0$, the integrand in Eq.~\eqref{eq:F}, its integral, and $y-y_0$ are all positive;
therefore $\widetilde{\mathcal{B}}(y)>0$. Moreover,  since $\widetilde{\mathcal{B}}(y_0)=0$ and
$\widetilde{\mathcal{B}}(y)\to0$ as $y\to\infty$ (the integrand behaves as $x^{2-\beta}$,
whose integral converges for $\beta>3$, whereas the denominator grows linearly in $y$), the function $\widetilde{\mathcal{B}}(y)$ has a finite positive maximum $\widetilde{\mathcal{B}}_{\max}$ at some radius $r_\textrm{m}\in(r_0,\infty)$ at which $\mathcal{B}(y)$ has
the minimum $\mathcal{B}_{\min}$.  The exact condition excluding
all exterior roots is therefore
\begin{equation}
 \frac{2{\rm M}_{\rm DM}^{\rm tot}}{r_{\rm s}}\widetilde{\mathcal{B}}_{\max}<1\,.
 \label{eq:no-root}
\end{equation}
Note that equality happens when  $\mathcal{B}_{\min}=0$.  Hence the absence of exterior
roots is not unconditional for an arbitrarily compact matter distribution,
but follows in the dilute galactic regime with ${\rm M}_{\rm DM}^{\rm tot}/{r_{\rm s}}\ll1$.

To see explicitly how this is related to ${\rm M}_{\rm DM}^{\rm tot}/r_{\rm s}\ll1$, we note that
$0\leq1-y_0/x\leq1$ outside the horizon $r_0$, which implies $0\leq\widetilde{\mathcal{B}}(y)\leq
(3-\gamma)h_{\max}$, where $h_{\max}$ is the unique maximum of function $h$ at
$x_*^\alpha=(2-\gamma)/(\beta-2)$. More explicitly, we have
\begin{equation}
 \begin{aligned}
 h_{\max}&=\left[
 \frac{(2-\gamma)^{2-\gamma}(\beta-2)^{\beta-2}}
 {(\beta-\gamma)^{\beta-\gamma}}
 \right]^{1/\alpha}<1\,,\\
 \widetilde{\mathcal{B}}_{\max}&\leq(3-\gamma)h_{\max}\le3\,.
 \end{aligned}
 \label{eq:bound}
\end{equation}
Accordingly, a parameter-dependent sufficient condition to satisfy the inequality \eqref{eq:no-root} is
${\rm M}_{\rm DM}^{\rm tot}/r_{\rm s}<[2(3-\gamma)h_{\max}]^{-1}$. Therefore, the condition ${\rm M}_{\rm DM}^{\rm tot}/r_{\rm s}<1/6$ is sufficient to satisfy \eqref{eq:no-root} for the full parameter range. The
assumption of dilute halos, i.e., ${\rm M}_{\rm DM}^{\rm tot}/r_{\rm s}\ll1$, therefore keeps
$\mathcal{B}$ positive everywhere outside $r_0$. Restoring units, this small
quantity is the halo compactness $G{\rm M}_{\rm DM}^{\rm tot}/(c^2r_{\rm s})\ll1$.

Together, for dilute halos, we find that
\begin{equation}
 \begin{aligned}
 \mathcal{B}(y)&>1 &&(0<r<r_0)\,,\\
 \mathcal{B}(y_0)&=1
\,,\\
 \mathcal{B}(y)&>0 &&(r>r_0)\,,
 \end{aligned}
 \label{eq:summary}
\end{equation}
and the only zero of the radial metric function is at $r=r_0$.

Taking the exact Hernquist case $(\alpha,\beta,\gamma)=(1,4,1)$ as an example, we have 
\begin{equation}
 \widetilde{\mathcal{B}}(y)=\frac{y-y_0}{(1+y_0)(1+y)^2}\,,
 \end{equation}
 whose maximum is
 \begin{equation}
\widetilde{\mathcal{B}}_{\max}=\frac{1}{4(1+y_0)^2}\,,
 \end{equation}
at $y=1+2y_0$. In this case, the exact no-root condition becomes ${\rm M}_{\rm DM}^{\rm tot}/r_{\rm s}<2(1+y_0)^2\simeq2$, which is
automatically satisfied when ${\rm M}_{\rm DM}^{\rm tot}/r_{\rm s}\ll1$.

\section{Photon sphere and QNM redshifts}\label{sec:phqnm}

BH QNMs, specifically their complex spectra, carry valuable information about the strong-field regimes near the BHs. In the presence of DM halos surrounding the BH, the gravity contributed by the halo shifts the QNM spectrum in a way that depends on the DM property and on how one models the halo systems. In Refs.~\cite{Pezzella:2024tkf,Bhowmik:2026owi}, by considering a spherically symmetric halo and assuming a zero radial pressure in the halo, i.e., $w=0$, it was found that the QNM spectra are redshifted in a way that the shifts on the real part and the imaginary part, relative to the Schwarzschild values, are equal to each other. More explicitly, the QNMs are redshifted as
\begin{equation}
\frac{\omega_R}{\omega_{\textrm{Sch},R}}\approx\frac{\omega_I}{\omega_{\textrm{Sch},I}}\approx1-\frac{C{\rm M}_{\rm DM}^{\rm tot}}{r_{\rm s}}+\mathcal{O}\left(\frac{{\rm M}_{\rm DM}^{\rm tot}}{r_{\rm s}}\right)^2\,,\label{qnmrelation}
\end{equation}
where $C>0$ depends on the DM profiles under consideration{\footnote{A similar redshift relation was also found in the system of a Schwarzschild BH encircled by a gravitating thin disk, which makes the system not spherically symmetric \cite{Chen:2023akf}.}}. On the plane spanned by the horizontal axis and the vertical axis defined by $\omega_R/\omega_{\textrm{Sch},R}$ and $\omega_I/\omega_{\textrm{Sch},I}$, respectively, the relation \eqref{qnmrelation} implies that, up to the first order in ${\rm M}_{\rm DM}^{\rm tot}/r_{\rm s}$, the QNM frequencies are redshifted from the point $(1,1)$, i.e., the pure Schwarzschild values, following a straight line with slope one (see the black line in Fig.~\ref{fig:dmqnm}).

The shifts of QNM frequencies can be understood using the eikonal correspondence, i.e., the well-known correspondence between the high-frequency QNMs and unstable light rings around the BH \cite{Cardoso:2008bp}. In the high-frequency limit, the QNM frequencies of high-$\ell$ QNMs can be approximated as
\begin{equation}
\omega_R\approx\ell\Omega_{\rm LR}\,,\qquad \omega_I\approx-\frac{1}{2}|\lambda_{\rm LR}|\,,
\end{equation}
where $\ell$ is the multipole number of QNMs, and $\Omega_{\rm LR}$ and $\lambda_{\rm LR}$ are the angular frequency and the Lyapunov exponent evaluated on the unstable light ring. Note that we only consider fundamental modes. In the setup considered in Refs.~\cite{Pezzella:2024tkf,Bhowmik:2026owi} (see also \cite{Hassanabadi:2026mmi}), one can explicitly show that 
\begin{equation}
\Omega_{\rm{LR}}\approx\lambda_{\rm{LR}}\approx\frac{2}{3\sqrt{3}r_0}\left(1-\frac{C{\rm M}_{\rm DM}^{\rm tot}}{r_{\rm s}}\right)\,,\label{lrshift}
\end{equation}
up to the first order in ${\rm M}_{\rm DM}^{\rm tot}/r_{\rm s}$.

The purpose of this section is to see how such a universal redshift relation would depend on different values of the equation of state $w$. In particular, we will mainly focus on the halo corrections induced on the light ring angular frequency $\Omega_{\rm{LR}}$ and the Lyapunov exponent $\lambda_{\rm{LR}}$. In the setups where $\Omega_{\rm{LR}}$ and $\lambda_{\rm{LR}}$ are not shifted in a way like Eq.~\eqref{lrshift}, we consider them the violation of such a universal QNM redshift relation. On the other hand, in the cases where $\Omega_{\rm{LR}}$ and $\lambda_{\rm{LR}}$ can be recast as in Eq.~\eqref{lrshift}, the universal redshift relation is justified for high-$\ell$ modes. For low values of $\ell$, we will directly compute the QNM frequencies numerically, and it turns out that the redshift relation \eqref{qnmrelation} is pretty valid already even for these low-$\ell$ modes.

\subsection{General cases: $w\ne-1$}\label{subsec:wnm1}
We first start with the general case with $ w\ne-1$. The photon sphere radius, or light ring, $r_{\rm LR}$ satisfies
\begin{equation}
2A(r_{\rm LR})-r_{\rm LR} A'(r_{\rm LR})=0\,.\label{lightringgeneral}
\end{equation}
At the leading order of ${\rm M}_{\rm DM}^{\rm tot}/r_{\rm s}$, the light ring radius is given by
\begin{equation}
r_{\rm LR}=\frac{3}{2}r_0\left(1-\frac{{\rm M}_{\rm DM}^{\rm tot}}{2r_{\rm s}}y_0\widetilde{\mathcal{A}}_{,y}(3y_0/2)+\mathcal{O}({\rm M}_{\rm DM}^{\rm tot}/r_{\rm s})^2\right)\,.
\end{equation}
The second term in the bracket should be handled with care because the behavior of $\widetilde{\mathcal{A}}_{,y}(3y_0/2)$ may differ significantly depending on the choice of $(\alpha,\beta,\gamma)$. We find it convenient to consider $y_0 \widetilde{\mathcal{A}}_{,y}(3y_0/2)$, which can be generically expressed as
\begin{align}
&y_0 \widetilde{\mathcal{A}}_{,y}\left(3y_0/2\right)\nonumber\\&\approx\frac{2(2-\gamma)}{3}\left(\frac{3}{2}y_0\right)^{2-\gamma}\left[1-\frac{(1+w)\Gamma\left(\frac{2-\gamma}{\alpha}\right)\Gamma\left(\frac{3+\alpha-\gamma}{\alpha}\right)}{\Gamma\left(\frac{3-\gamma}{\alpha}\right)\Gamma\left(\frac{2+\alpha-\gamma}{\alpha}\right)}\right]\,.
\end{align}
Therefore, the light ring radius becomes
\begin{widetext}
\begin{equation}
r_{\rm LR}=\frac{3}{2}r_0\left\{1-\frac{{\rm M}_{\rm DM}^{\rm tot}(2-\gamma)}{3r_{\rm s}}\left(\frac{3r_0}{2r_{\rm s}}\right)^{2-\gamma}\left[1-\frac{(1+w)\Gamma\left(\frac{2-\gamma}{\alpha}\right)\Gamma\left(\frac{3+\alpha-\gamma}{\alpha}\right)}{\Gamma\left(\frac{3-\gamma}{\alpha}\right)\Gamma\left(\frac{2+\alpha-\gamma}{\alpha}\right)}\right]+\mathcal{O}({\rm M}_{\rm DM}^{\rm tot}/r_{\rm s})^2\right\}\,.
\end{equation}
\end{widetext}
This reproduces the result of the light ring shifts shown in Ref.~\cite{Cardoso:2021wlq} when $(\alpha,\beta,\gamma)=(1,4,1)$ and $w=0$.

The angular frequency on the light ring is given by
\begin{align}
\Omega_{\textrm{LR}}&\equiv\frac{\sqrt{A(r_{\rm LR})}}{r_{\rm LR}}\nonumber\\
&\approx\frac{2}{3\sqrt{3}r_0}\left(1-\frac{{\rm M}_{\rm DM}^{\rm tot}}{r_{\rm s}}\widetilde{\mathcal{A}}(0)\right)\,.\label{olra0}
\end{align}
Furthermore, the Lyapunov exponent on the light ring is defined as
\begin{equation}
\lambda_{\textrm{LR}}\equiv\sqrt{\left(1-\frac{2M(r_{\rm LR})}{r_{\rm LR}}\right)\frac{\left(2A(r_{\rm LR})-r_{\rm LR}^2A''(r_{\rm LR})\right)}{2r_{\rm LR}^2}}\,.\label{lyapunovgeneral}
\end{equation}
At the leading order of ${\rm M}_{\rm DM}^{\rm tot}/r_{\rm s}$, we get
\begin{equation}
\lambda_{\textrm{LR}}\approx\frac{2}{3\sqrt{3}r_0}\left(1-\frac{{\rm M}_{\rm DM}^{\rm tot}}{r_{\rm s}}\widetilde{\mathcal{A}}(0)\right)\,.\label{lambdalra0}
\end{equation}
One can see from Eqs.~\eqref{olra0} and \eqref{lambdalra0} that the angular frequency and the Lyapunov exponent on the light ring can be expressed as those in Eq.~\eqref{lrshift}. Note that in this case, we have $\widetilde{\mathcal{A}}(0)=-(w+1)H(0)$ and $H(0)$ is given by Eq.~\eqref{At0general}. Therefore, the eikonal correspondence suggests that in this case, the real part and the imaginary part of QNMs are redshifted by the same amount as long as $w\ne -1$, and the universal QNM redshift relation may hold (see Fig.~\ref{fig:dmqnm} for the numerical results of QNM spectra).

We would like to mention that our analysis here is consistent with the result of Ref.~\cite{Ylla:2026ffv}, which discusses the shifts of $\Omega_{\textrm{LR}}$ and $\lambda_{\textrm{LR}}$ of hairy BHs, whose hair is modeled by anisotropic fluids minimally coupled to general relativity. At the first-order deformations generated by the fluid, the difference between the shifts of $\Omega_{\textrm{LR}}$ and $\lambda_{\textrm{LR}}$ is controlled by $M'(r)(1+w_t)$ evaluated at the Schwarzschild light ring, where $w_t$ is defined by $w_t\equiv p_t/\rho$ (see Eq. (2.31) of Ref.~\cite{Ylla:2026ffv}). From Eqs.~\eqref{mbpara} and \eqref{bybtilde} and taking $y_0\ll1$, one can see that $M'(0)\approx {\rm M}_{\rm DM}^{\rm tot}\widetilde{\mathcal{B}}(0)/r_{\rm s}=0$, implying that in the presence of fluids, $\Omega_{\textrm{LR}}$ and $\lambda_{\textrm{LR}}$ are shifted by the same amount, as long as this common shift is not suppressed.

\subsection{Special case: $w=-1$}\label{subsec:wm1}
In the special case where $w=-1$, one cannot follow the previous procedures to obtain the corrections on the light ring quantities. The big difference is that $\widetilde{\mathcal{A}}(0)$, which appears as the leading-order corrections on the light ring quantities, is given by $\widetilde{\mathcal{A}}(0)=\widetilde{\mathcal{B}}(0)$ and would vanish in this case (see Eq.~\eqref{Bwidetildegen}). Therefore, one has to take into account the contributions from several $y_0$-associated terms that have been neglected in the derivation of Eq.~\eqref{Bwidetildegen} in the previous section. In this case, the corrections on all light ring quantities are further suppressed by the factor $y_0^{2-\gamma}$, and should be discussed separately from the general case $w\ne -1$.

\begin{figure*}
  \centering
  \subfigure[]
{ \includegraphics[scale=0.37]{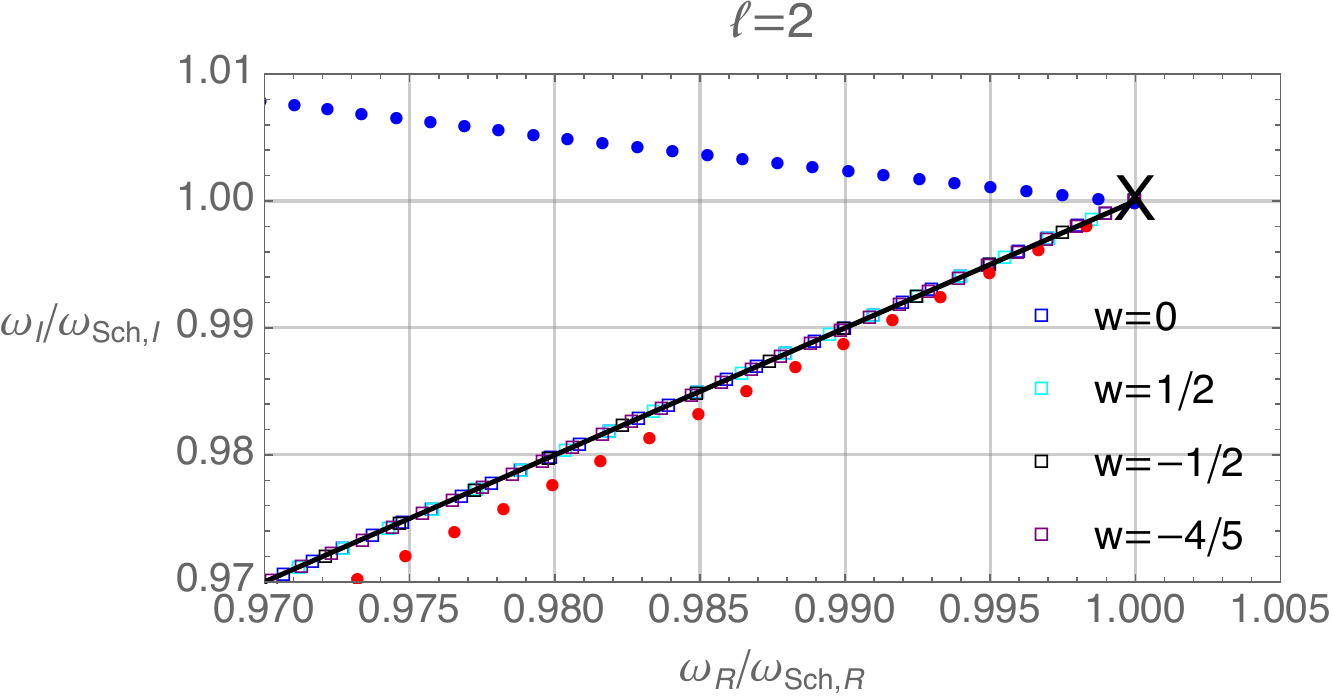}}
\subfigure[]
  {\includegraphics[scale=0.37]{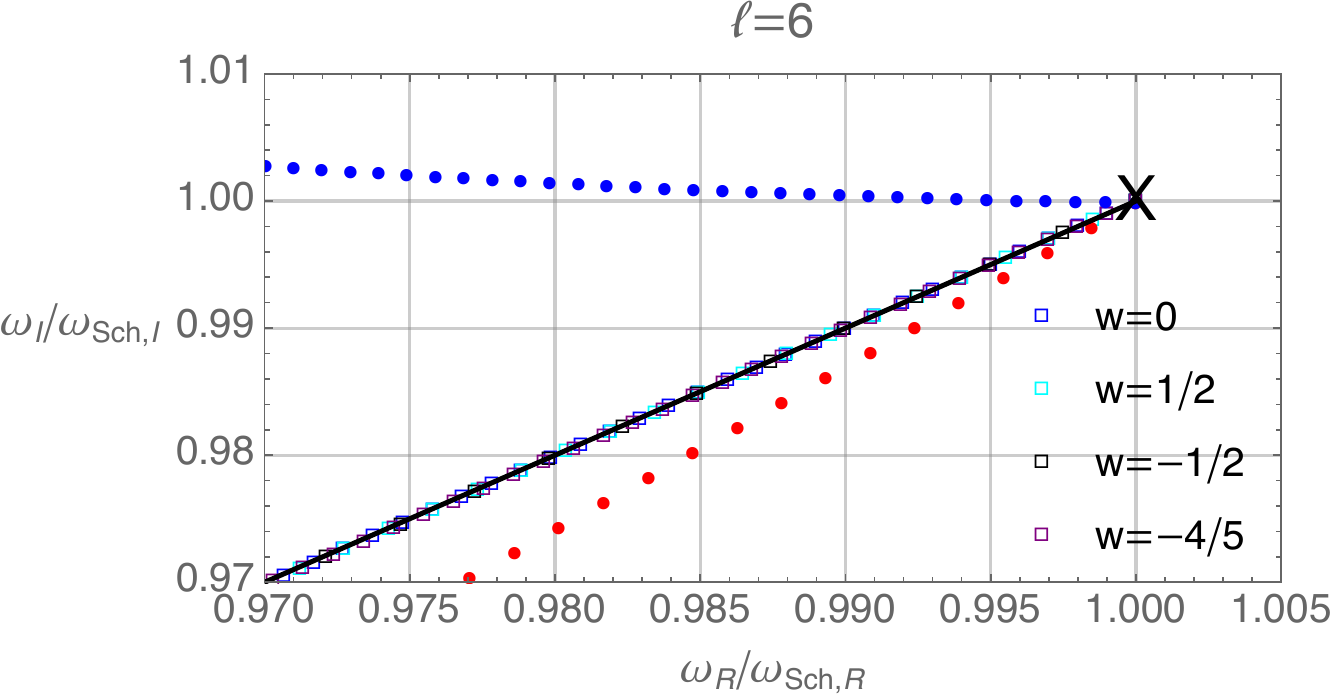}}
\\
\caption{The fundamental QNM frequency of massless scalar fields in the presence of a DM halo, with (a) $\ell=2$ and (b) $\ell=6$. We fix $r_{\rm s}=1000M_{\rm BH}$. The horizontal and vertical axes show the real and the imaginary parts of the frequencies relative to the Schwarzschild values. The cross on the right end in each panel represents the Schwarzschild case, i.e., $(1,1)$, from which we increase the value of ${\rm M}_{\rm DM}^{\rm tot}/r_{\rm s}$ for each $w$ and compute the QNM frequencies. Here we choose the Hernquist-type model with $(\alpha,\beta,\gamma)=(1,4,1)$. For $w\ne-1$, the QNM spectra are universally redshifted following a straight line with slope one, i.e., the black straight line. For the special case $w=-1$, the red and blue circles represent the results obtained by using the approximated formula \eqref{Bwidetildegen} for the metric and the exact mass function \eqref{exactmass}, respectively.}
\label{fig:dmqnm} 
\end{figure*}

In the special case where $w=-1$, the redshift function is
\begin{equation}
A(r)=1-\frac{2M(r)}{r}\,,\label{Aspecial1}
\end{equation}
with mass function given by Eq.~\eqref{exactmass}. It should be emphasized that here we use the exact mass function \eqref{exactmass} rather than the approximation \eqref{Bwidetildegen}. In the derivation of Eq.~\eqref{Bwidetildegen}, some terms that contain $y_0^p$ with power $p>0$ are already neglected. However, in the special case of $w=-1$, the leading-order corrections in the light ring quantities will be associated with terms that have been neglected (see Eq.~\eqref{qgamma}). Therefore, directly using Eq.~\eqref{Bwidetildegen} in this case gives rise to inconsistent results, which differ significantly from those obtained using the exact formula \eqref{exactmass} (see the blue and red circles in Fig.~\ref{fig:dmqnm}).

For convenience, we define the effective halo compactness \cite{Hassanabadi:2026wgc}
\begin{equation}
q_\gamma\equiv\frac{{\rm M}_{\rm DM}^{\rm tot}}{r_{\rm s}}\left(\frac{r_0}{r_{\rm s}}\right)^{2-\gamma}\,,\label{qgamma}
\end{equation}
which perturbatively quantifies the halo corrections. Using Eq.~\eqref{lightringgeneral}, the light ring radius, up to the leading order in $q_\gamma$, is given by 
\begin{equation}
r_{\rm LR}\approx\frac{3}{2}r_0\left[1-q_\gamma\frac{3^{\gamma-1}2-2^{\gamma-2}\left(\gamma^2-2\gamma+6\right)}{3^{\gamma-1}(\gamma-2)}\right]\,.
\end{equation}
Then, the angular frequency $\Omega_{\textrm{LR}}$ can be approximated as
\begin{equation}
\Omega_{\textrm{LR}}\approx\frac{2}{3\sqrt{3}r_0}\left[1+\frac{q_\gamma\left(2-\left(\frac{2}{3}\right)^{\gamma-2}\gamma\right)}{\gamma-2}\right]\,.\label{omegalrm1}
\end{equation}
Finally, the Lyapunov exponent \eqref{lyapunovgeneral} reads
\begin{align}
\lambda_{\textrm{LR}}\approx&\,\frac{2}{3\sqrt{3}r_0}\nonumber\\&\times\left[1+q_\gamma\frac{2^{\gamma-2}\left(\gamma^3-7\gamma^2+10\gamma-12\right)+3^{\gamma-1}4}{3^{\gamma-1}(2\gamma-4)}\right]\,,\label{lambdalrm1}
\end{align}
up to the same order in $q_\gamma$. One can see that up to the first order in $q_\gamma$, the corrections on $\Omega_{\textrm{LR}}$ and $\lambda_{\textrm{LR}}$ appear in a way such that the relation \eqref{lrshift} fails to apply. Therefore, the universal QNM redshift relation \eqref{qnmrelation} is not valid either.

In fact, one also sees that the corrections on $\Omega_{\textrm{LR}}$ and $\lambda_{\textrm{LR}}$ only depend on the inner slope $\gamma$. For instance, when $\gamma=1$, the angular frequency is redshifted as
\begin{equation}
\Omega_{\textrm{LR}}\approx\frac{2}{3\sqrt{3}r_0}\left(1-\frac{q_1}{2}\right)\,,
\end{equation}
while the Lyapunov exponent does not receive corrections at this order of $q_\gamma$. It should be emphasized that the redshift formulas for $\Omega_{\textrm{LR}}$ and $\lambda_{\textrm{LR}}$ obtained in Eqs.~\eqref{omegalrm1} and \eqref{lambdalrm1} are different from those obtained in Ref.~\cite{Hassanabadi:2026wgc} (Eqs.~(56) and (58)). This is because in Ref.~\cite{Hassanabadi:2026wgc} the factor $b(r)=1-r_0/r$ was not included in the DM energy density to force $\rho$ to vanish at the horizon. Since the redshift formulas in the case of $w=-1$ are sensitive to the inner structure of the DM density profile, which highly depends on the existence of the extra factor $b(r)$, it is thus not surprising that adding $b(r)$ can drastically modify the redshift formula in this special case $w=-1$. 

In Fig.~\ref{fig:dmqnm}, we compute the fundamental QNM frequency of massless scalar fields in the presence of a DM halo with $r_{\rm s}=1000M_{\rm BH}$ relative to the pure Schwarzschild QNM frequency. The horizontal and the vertical axes represent the real and the imaginary parts of the spectra, respectively. On the left (right) panel, we take $\ell=2$ ($\ell=6$). We focus on the Hernquist-type model with $(\alpha,\beta,\gamma)=(1,4,1)$, and the open squared points with different colors represent different values of $w$. For each $w$, we increase the value of ${\rm M}_{\rm DM}^{\rm tot}/r_{\rm s}$ from zero, i.e., the pure Schwarzschild case at $(1,1)$. One sees that as long as $w\ne-1$, the QNMs are redshifted following a straight line with slope one, i.e., the black straight line. However, if $w=-1$, the QNM spectra are shifted in a way depending on the value of $\gamma$. In Fig.~\ref{fig:dmqnm}, the red and the blue circles represent the QNM results of $w=-1$ evaluated using the approximated formula \eqref{Bwidetildegen} for the metric and the exact mass function \eqref{exactmass}, respectively. In particular, when $\ell$ increases, the first few blue circles distribute almost horizontally to the left of the Schwarzschild value $(1,1)$. This is consistent with the redshift formulas \eqref{omegalrm1} and \eqref{lambdalrm1} with $\gamma=1$. 

Finally, we would like to mention that although in our setup we only consider $\beta>3$ to ensure the total DM mass to be finite, one can relax this assumption on $\beta$ while still maintaining a finite DM mass by introducing a cutoff length parameter $s$ as the radius of the halo such that $s>r_{\rm s}\gg r_0$ and $\rho(r>s)=0$ \cite{Konoplya:2022hbl}. We expect that our results in terms of the relation between QNM redshifts and the equation of state $w$ should still be robust in these cases so long as $y_0\equiv r_0/r_{\rm s}\ll1$ is justified. When $y_0\ll1$, the function $\widetilde{\mathcal{B}}(y)$ still approaches zero when $y\rightarrow0$, meaning that the redshift factor $\widetilde{\mathcal{A}}(0)$ is still given by $\widetilde{\mathcal{A}}(0)=-(1+w)H(0)$. In Ref.~\cite{Konoplya:2022hbl} it was shown that $H(0)$ depends on $s$ and $\beta$, but it in general does not vanish unless $s\rightarrow\infty$, i.e., a halo with divergent mass. As a result, if $w\ne-1$ such that $\widetilde{\mathcal{A}}(0)\ne0$, the QNM redshifts still follow the universal pattern, even for $\beta\le3$. On the other hand, for $w=-1$, we have $\widetilde{\mathcal{A}}(0)=0$, and the universal redshift relation is broken.

\section{Conclusions}\label{sec:conclusion}

The program of probing strong gravity or testing general relativity through BH spectroscopy, i.e., via BH QNMs, is based on the fact that QNM spectra of BHs only carry information about the BH spacetime itself. However, once the gravitational backreaction of the surrounding environments, i.e., DM halo, galactic structure, accretion disk, companions, etc, are taken into account, tiny modifications of QNM spectra would appear. It is therefore crucial to break the degeneracy between such tiny environmental effects and the modifications arising from putative new physics beyond general relativity, especially given that the contributions from the latter are also expected to be small. The universal QNM redshift relation induced by a dilute and spherically symmetric halo with a zero radial pressure \cite{Cardoso:2021wlq,Pezzella:2024tkf,Bhowmik:2026owi}, i.e., Eq.~\eqref{qnmrelation}, seems to be a candidate to break such degeneracy. In particular, at the leading order of the halo compactness ${\rm M}_{\rm DM}^{\rm tot}/r_{\rm s}$, the QNM spectra are redshifted on the complex plane in a universal manner, i.e., the black straight lines in Fig.~\ref{fig:dmqnm}.

In the present paper, we extend the analysis in Refs.~\cite{Cardoso:2021wlq,Pezzella:2024tkf,Bhowmik:2026owi} by relaxing the assumption of zero radial pressure. By modeling the BH-halo system as a dilute gravitating configuration supported by an anisotropic fluid with an effective radial equation of state, i.e., $p_r=w\rho$, we find that the universal QNM redshift relation holds as long as $w\ne-1$. The leading-order corrections to the light ring frequency, Lyapunov exponent, and the QNM redshifts are controlled by a factor proportional to $w+1$, which vanishes exactly when $w=-1$. In this special case, the leading-order QNM shifts depend not only on the compactness, but also on the inner structure of the halo density, and the universal redshift relation is not valid. 

The results in this paper imply the validity of the universal QNM redshift relation \eqref{qnmrelation} beyond the setup restricted to zero radial pressure. It is valid for a wider class of DM halo models potentially including Self-Interacting Dark Matter models, Fuzzy Dark Matter models, and other phenomenological setups. 

A natural extension of this work is to go beyond spherical symmetry. In fact, based on the assumption that the mass of the accretion disk is much smaller than the BH mass, the QNM spectra of a BH surrounded by a static and gravitating thin disk seem to follow a similar redshift relation \cite{Chen:2023akf}. In the spherically symmetric configuration, the analytic treatment showing such a universal relation can be carried out straightforwardly through the eikonal correspondence that connects the QNM redshifts to the shifts of light ring frequency and Lyapunov exponent. However, in the absence of spherical symmetry, the light ring trajectories are non-trivially deformed, the radial and polar-angle sectors of the geodesic equations cannot be separable, and QNMs split into different azimuthal modes. All these ingredients largely complicate the analytic treatment of the universal redshift relation. It remains to be explored whether the robustness of the universal redshift relation extends beyond spherical symmetry. 

\acknowledgments
 C.Y.C is supported by the Special Postdoctoral
Researcher (SPDR) Program at RIKEN and RIKEN
Incentive Research Grant (Shoreikadai) 2025. The author would like to thank Vitor Cardoso for several valuable and insightful discussions during the preparation of this work.

\bibliographystyle{mybibstyle}
\bibliography{bib}

\begin{thebibliography}{86}%
\makeatletter
\providecommand \@ifxundefined [1]{%
 \@ifx{#1\undefined}
}%
\providecommand \@ifnum [1]{%
 \ifnum #1\expandafter \@firstoftwo
 \else \expandafter \@secondoftwo
 \fi
}%
\providecommand \@ifx [1]{%
 \ifx #1\expandafter \@firstoftwo
 \else \expandafter \@secondoftwo
 \fi
}%
\providecommand \natexlab [1]{#1}%
\providecommand \enquote  [1]{``#1''}%
\providecommand \bibnamefont  [1]{#1}%
\providecommand \bibfnamefont [1]{#1}%
\providecommand \citenamefont [1]{#1}%
\providecommand \href@noop [0]{\@secondoftwo}%
\providecommand \href [0]{\begingroup \@sanitize@url \@href}%
\providecommand \@href[1]{\@@startlink{#1}\@@href}%
\providecommand \@@href[1]{\endgroup#1\@@endlink}%
\providecommand \@sanitize@url [0]{\catcode `\\12\catcode `\$12\catcode
  `\&12\catcode `\#12\catcode `\^12\catcode `\_12\catcode `\%12\relax}%
\providecommand \@@startlink[1]{}%
\providecommand \@@endlink[0]{}%
\providecommand \url  [0]{\begingroup\@sanitize@url \@url }%
\providecommand \@url [1]{\endgroup\@href {#1}{\urlprefix }}%
\providecommand \urlprefix  [0]{URL }%
\providecommand \Eprint [0]{\href }%
\providecommand \doibase [0]{http://dx.doi.org/}%
\providecommand \selectlanguage [0]{\@gobble}%
\providecommand \bibinfo  [0]{\@secondoftwo}%
\providecommand \bibfield  [0]{\@secondoftwo}%
\providecommand \translation [1]{[#1]}%
\providecommand \BibitemOpen [0]{}%
\providecommand \bibitemStop [0]{}%
\providecommand \bibitemNoStop [0]{.\EOS\space}%
\providecommand \EOS [0]{\spacefactor3000\relax}%
\providecommand \BibitemShut  [1]{\csname bibitem#1\endcsname}%
\let\auto@bib@innerbib\@empty
%</preamble>
\bibitem [{\citenamefont {Abbott}\ \emph {et~al.}(2016)\citenamefont {Abbott}
  \emph {et~al.}}]{LIGOScientific:2016aoc}%
  \BibitemOpen
  \bibfield  {author} {\bibinfo {author} {\bibfnamefont {B.~P.}\ \bibnamefont
  {Abbott}} \emph {et~al.} (\bibinfo {collaboration} {LIGO Scientific,
  Virgo}),\ }\href {\doibase 10.1103/PhysRevLett.116.061102} {\bibfield
  {journal} {\bibinfo  {journal} {\emph {Phys. Rev. Lett.}}\ }\textbf {\bibinfo
  {volume} {116}},\ \bibinfo {pages} {061102} (\bibinfo {year} {2016})},\
  \Eprint {http://arxiv.org/abs/1602.03837} {arXiv:1602.03837 [gr-qc]}
  \BibitemShut {NoStop}%
\bibitem [{\citenamefont {Abbott}\ \emph {et~al.}(2021)\citenamefont {Abbott}
  \emph {et~al.}}]{LIGOScientific:2020ibl}%
  \BibitemOpen
  \bibfield  {author} {\bibinfo {author} {\bibfnamefont {R.}~\bibnamefont
  {Abbott}} \emph {et~al.} (\bibinfo {collaboration} {LIGO Scientific,
  Virgo}),\ }\href {\doibase 10.1103/PhysRevX.11.021053} {\bibfield  {journal}
  {\bibinfo  {journal} {\emph {Phys. Rev. X}}\ }\textbf {\bibinfo {volume}
  {11}},\ \bibinfo {pages} {021053} (\bibinfo {year} {2021})},\ \Eprint
  {http://arxiv.org/abs/2010.14527} {arXiv:2010.14527 [gr-qc]} \BibitemShut
  {NoStop}%
\bibitem [{\citenamefont {Abbott}\ \emph {et~al.}(2023)\citenamefont {Abbott}
  \emph {et~al.}}]{KAGRA:2021vkt}%
  \BibitemOpen
  \bibfield  {author} {\bibinfo {author} {\bibfnamefont {R.}~\bibnamefont
  {Abbott}} \emph {et~al.} (\bibinfo {collaboration} {KAGRA, VIRGO, LIGO
  Scientific}),\ }\href {\doibase 10.1103/PhysRevX.13.041039} {\bibfield
  {journal} {\bibinfo  {journal} {\emph {Phys. Rev. X}}\ }\textbf {\bibinfo
  {volume} {13}},\ \bibinfo {pages} {041039} (\bibinfo {year} {2023})},\
  \Eprint {http://arxiv.org/abs/2111.03606} {arXiv:2111.03606 [gr-qc]}
  \BibitemShut {NoStop}%
\bibitem [{\citenamefont {Davis}\ \emph {et~al.}(2021)\citenamefont {Davis}
  \emph {et~al.}}]{LIGO:2021ppb}%
  \BibitemOpen
  \bibfield  {author} {\bibinfo {author} {\bibfnamefont {D.}~\bibnamefont
  {Davis}} \emph {et~al.} (\bibinfo {collaboration} {LIGO}),\ }\href {\doibase
  10.1088/1361-6382/abfd85} {\bibfield  {journal} {\bibinfo  {journal} {\emph
  {Class. Quant. Grav.}}\ }\textbf {\bibinfo {volume} {38}},\ \bibinfo {pages}
  {135014} (\bibinfo {year} {2021})},\ \Eprint
  {http://arxiv.org/abs/2101.11673} {arXiv:2101.11673 [astro-ph.IM]}
  \BibitemShut {NoStop}%
\bibitem [{\citenamefont {Ghez}\ \emph {et~al.}(1998)\citenamefont {Ghez},
  \citenamefont {Klein}, \citenamefont {Morris},\ and\ \citenamefont
  {Becklin}}]{Ghez:1998ph}%
  \BibitemOpen
  \bibfield  {author} {\bibinfo {author} {\bibfnamefont {A.~M.}\ \bibnamefont
  {Ghez}}, \bibinfo {author} {\bibfnamefont {B.~L.}\ \bibnamefont {Klein}},
  \bibinfo {author} {\bibfnamefont {M.}~\bibnamefont {Morris}},  and \bibinfo
  {author} {\bibfnamefont {E.~E.}\ \bibnamefont {Becklin}},\ }\href {\doibase
  10.1086/306528} {\bibfield  {journal} {\bibinfo  {journal} {\emph {Astrophys.
  J.}}\ }\textbf {\bibinfo {volume} {509}},\ \bibinfo {pages} {678} (\bibinfo
  {year} {1998})},\ \Eprint {http://arxiv.org/abs/astro-ph/9807210}
  {arXiv:astro-ph/9807210} \BibitemShut {NoStop}%
\bibitem [{\citenamefont {Genzel}\ \emph {et~al.}(2010)\citenamefont {Genzel},
  \citenamefont {Eisenhauer},\ and\ \citenamefont
  {Gillessen}}]{genzel2010galactic}%
  \BibitemOpen
  \bibfield  {author} {\bibinfo {author} {\bibfnamefont {R.}~\bibnamefont
  {Genzel}}, \bibinfo {author} {\bibfnamefont {F.}~\bibnamefont {Eisenhauer}},
  and \bibinfo {author} {\bibfnamefont {S.}~\bibnamefont {Gillessen}},\
  }\href@noop {} {\bibfield  {journal} {\bibinfo  {journal} {\emph {Reviews of
  modern physics}}\ }\textbf {\bibinfo {volume} {82}},\ \bibinfo {pages} {3121}
  (\bibinfo {year} {2010})}\BibitemShut {NoStop}%
\bibitem [{\citenamefont {Akiyama}\ \emph {et~al.}(2019)\citenamefont {Akiyama}
  \emph {et~al.}}]{EventHorizonTelescope:2019dse}%
  \BibitemOpen
  \bibfield  {author} {\bibinfo {author} {\bibfnamefont {K.}~\bibnamefont
  {Akiyama}} \emph {et~al.} (\bibinfo {collaboration} {Event Horizon
  Telescope}),\ }\href {\doibase 10.3847/2041-8213/ab0ec7} {\bibfield
  {journal} {\bibinfo  {journal} {\emph {Astrophys. J. Lett.}}\ }\textbf
  {\bibinfo {volume} {875}},\ \bibinfo {pages} {L1} (\bibinfo {year} {2019})},\
  \Eprint {http://arxiv.org/abs/1906.11238} {arXiv:1906.11238 [astro-ph.GA]}
  \BibitemShut {NoStop}%
\bibitem [{\citenamefont {Akiyama}\ \emph {et~al.}(2022)\citenamefont {Akiyama}
  \emph {et~al.}}]{EventHorizonTelescope:2022wkp}%
  \BibitemOpen
  \bibfield  {author} {\bibinfo {author} {\bibfnamefont {K.}~\bibnamefont
  {Akiyama}} \emph {et~al.} (\bibinfo {collaboration} {Event Horizon
  Telescope}),\ }\href {\doibase 10.3847/2041-8213/ac6674} {\bibfield
  {journal} {\bibinfo  {journal} {\emph {Astrophys. J. Lett.}}\ }\textbf
  {\bibinfo {volume} {930}},\ \bibinfo {pages} {L12} (\bibinfo {year}
  {2022})},\ \Eprint {http://arxiv.org/abs/2311.08680} {arXiv:2311.08680
  [astro-ph.HE]} \BibitemShut {NoStop}%
\bibitem [{\citenamefont {Kokkotas}\ and\ \citenamefont
  {Schmidt}(1999)}]{Kokkotas:1999bd}%
  \BibitemOpen
  \bibfield  {author} {\bibinfo {author} {\bibfnamefont {K.~D.}\ \bibnamefont
  {Kokkotas}} and \bibinfo {author} {\bibfnamefont {B.~G.}\ \bibnamefont
  {Schmidt}},\ }\href {\doibase 10.12942/lrr-1999-2} {\bibfield  {journal}
  {\bibinfo  {journal} {\emph {Living Rev. Rel.}}\ }\textbf {\bibinfo {volume}
  {2}},\ \bibinfo {pages} {2} (\bibinfo {year} {1999})},\ \Eprint
  {http://arxiv.org/abs/gr-qc/9909058} {arXiv:gr-qc/9909058} \BibitemShut
  {NoStop}%
\bibitem [{\citenamefont {Berti}\ \emph {et~al.}(2009)\citenamefont {Berti},
  \citenamefont {Cardoso},\ and\ \citenamefont {Starinets}}]{Berti:2009kk}%
  \BibitemOpen
  \bibfield  {author} {\bibinfo {author} {\bibfnamefont {E.}~\bibnamefont
  {Berti}}, \bibinfo {author} {\bibfnamefont {V.}~\bibnamefont {Cardoso}},  and
  \bibinfo {author} {\bibfnamefont {A.~O.}\ \bibnamefont {Starinets}},\ }\href
  {\doibase 10.1088/0264-9381/26/16/163001} {\bibfield  {journal} {\bibinfo
  {journal} {\emph {Class. Quant. Grav.}}\ }\textbf {\bibinfo {volume} {26}},\
  \bibinfo {pages} {163001} (\bibinfo {year} {2009})},\ \Eprint
  {http://arxiv.org/abs/0905.2975} {arXiv:0905.2975 [gr-qc]} \BibitemShut
  {NoStop}%
\bibitem [{\citenamefont {Konoplya}\ and\ \citenamefont
  {Zhidenko}(2011)}]{Konoplya:2011qq}%
  \BibitemOpen
  \bibfield  {author} {\bibinfo {author} {\bibfnamefont {R.~A.}\ \bibnamefont
  {Konoplya}} and \bibinfo {author} {\bibfnamefont {A.}~\bibnamefont
  {Zhidenko}},\ }\href {\doibase 10.1103/RevModPhys.83.793} {\bibfield
  {journal} {\bibinfo  {journal} {\emph {Rev. Mod. Phys.}}\ }\textbf {\bibinfo
  {volume} {83}},\ \bibinfo {pages} {793} (\bibinfo {year} {2011})},\ \Eprint
  {http://arxiv.org/abs/1102.4014} {arXiv:1102.4014 [gr-qc]} \BibitemShut
  {NoStop}%
\bibitem [{\citenamefont
  {Vishveshwara}(1970{\natexlab{a}})}]{Vishveshwara:1970zz}%
  \BibitemOpen
  \bibfield  {author} {\bibinfo {author} {\bibfnamefont {C.~V.}\ \bibnamefont
  {Vishveshwara}},\ }\href {\doibase 10.1038/227936a0} {\bibfield  {journal}
  {\bibinfo  {journal} {\emph {Nature}}\ }\textbf {\bibinfo {volume} {227}},\
  \bibinfo {pages} {936} (\bibinfo {year} {1970}{\natexlab{a}})}\BibitemShut
  {NoStop}%
\bibitem [{\citenamefont {Berti}\ \emph {et~al.}(2006)\citenamefont {Berti},
  \citenamefont {Cardoso},\ and\ \citenamefont {Will}}]{Berti:2005ys}%
  \BibitemOpen
  \bibfield  {author} {\bibinfo {author} {\bibfnamefont {E.}~\bibnamefont
  {Berti}}, \bibinfo {author} {\bibfnamefont {V.}~\bibnamefont {Cardoso}},  and
  \bibinfo {author} {\bibfnamefont {C.~M.}\ \bibnamefont {Will}},\ }\href
  {\doibase 10.1103/PhysRevD.73.064030} {\bibfield  {journal} {\bibinfo
  {journal} {\emph {Phys. Rev. D}}\ }\textbf {\bibinfo {volume} {73}},\
  \bibinfo {pages} {064030} (\bibinfo {year} {2006})},\ \Eprint
  {http://arxiv.org/abs/gr-qc/0512160} {arXiv:gr-qc/0512160} \BibitemShut
  {NoStop}%
\bibitem [{\citenamefont {Giesler}\ \emph {et~al.}(2019)\citenamefont
  {Giesler}, \citenamefont {Isi}, \citenamefont {Scheel},\ and\ \citenamefont
  {Teukolsky}}]{Giesler:2019uxc}%
  \BibitemOpen
  \bibfield  {author} {\bibinfo {author} {\bibfnamefont {M.}~\bibnamefont
  {Giesler}}, \bibinfo {author} {\bibfnamefont {M.}~\bibnamefont {Isi}},
  \bibinfo {author} {\bibfnamefont {M.~A.}\ \bibnamefont {Scheel}},  and
  \bibinfo {author} {\bibfnamefont {S.}~\bibnamefont {Teukolsky}},\ }\href
  {\doibase 10.1103/PhysRevX.9.041060} {\bibfield  {journal} {\bibinfo
  {journal} {\emph {Phys. Rev. X}}\ }\textbf {\bibinfo {volume} {9}},\ \bibinfo
  {pages} {041060} (\bibinfo {year} {2019})},\ \Eprint
  {http://arxiv.org/abs/1903.08284} {arXiv:1903.08284 [gr-qc]} \BibitemShut
  {NoStop}%
\bibitem [{\citenamefont {Baibhav}\ \emph {et~al.}(2023)\citenamefont
  {Baibhav}, \citenamefont {Cheung}, \citenamefont {Berti}, \citenamefont
  {Cardoso}, \citenamefont {Carullo}, \citenamefont {Cotesta}, \citenamefont
  {Del~Pozzo},\ and\ \citenamefont {Duque}}]{Baibhav:2023clw}%
  \BibitemOpen
  \bibfield  {author} {\bibinfo {author} {\bibfnamefont {V.}~\bibnamefont
  {Baibhav}}, \bibinfo {author} {\bibfnamefont {M.~H.-Y.}\ \bibnamefont
  {Cheung}}, \bibinfo {author} {\bibfnamefont {E.}~\bibnamefont {Berti}},
  \bibinfo {author} {\bibfnamefont {V.}~\bibnamefont {Cardoso}}, \bibinfo
  {author} {\bibfnamefont {G.}~\bibnamefont {Carullo}}, \bibinfo {author}
  {\bibfnamefont {R.}~\bibnamefont {Cotesta}}, \bibinfo {author} {\bibfnamefont
  {W.}~\bibnamefont {Del~Pozzo}},  and \bibinfo {author} {\bibfnamefont
  {F.}~\bibnamefont {Duque}},\ }\href {\doibase 10.1103/PhysRevD.108.104020}
  {\bibfield  {journal} {\bibinfo  {journal} {\emph {Phys. Rev. D}}\ }\textbf
  {\bibinfo {volume} {108}},\ \bibinfo {pages} {104020} (\bibinfo {year}
  {2023})},\ \Eprint {http://arxiv.org/abs/2302.03050} {arXiv:2302.03050
  [gr-qc]} \BibitemShut {NoStop}%
\bibitem [{\citenamefont {Destounis}\ and\ \citenamefont
  {Duque}(2024)}]{destounis2024black}%
  \BibitemOpen
  \bibfield  {author} {\bibinfo {author} {\bibfnamefont {K.}~\bibnamefont
  {Destounis}} and \bibinfo {author} {\bibfnamefont {F.}~\bibnamefont
  {Duque}},\ }in\ \href@noop {} {\emph {\bibinfo {booktitle} {Compact objects
  in the universe}}}\ (\bibinfo  {publisher} {Springer},\ \bibinfo {year}
  {2024})\ pp.\ \bibinfo {pages} {155--202}\BibitemShut {NoStop}%
\bibitem [{\citenamefont {Cardoso}\ and\ \citenamefont
  {Pani}(2019)}]{Cardoso:2019rvt}%
  \BibitemOpen
  \bibfield  {author} {\bibinfo {author} {\bibfnamefont {V.}~\bibnamefont
  {Cardoso}} and \bibinfo {author} {\bibfnamefont {P.}~\bibnamefont {Pani}},\
  }\href {\doibase 10.1007/s41114-019-0020-4} {\bibfield  {journal} {\bibinfo
  {journal} {\emph {Living Rev. Rel.}}\ }\textbf {\bibinfo {volume} {22}},\
  \bibinfo {pages} {4} (\bibinfo {year} {2019})},\ \Eprint
  {http://arxiv.org/abs/1904.05363} {arXiv:1904.05363 [gr-qc]} \BibitemShut
  {NoStop}%
\bibitem [{\citenamefont {Berti}\ \emph {et~al.}(2026)\citenamefont {Berti}
  \emph {et~al.}}]{Berti:2025hly}%
  \BibitemOpen
  \bibfield  {author} {\bibinfo {author} {\bibfnamefont {E.}~\bibnamefont
  {Berti}} \emph {et~al.},\ }\href {\doibase 10.1088/1361-6382/ae59e2}
  {\bibfield  {journal} {\bibinfo  {journal} {\emph {Class. Quant. Grav.}}\
  }\textbf {\bibinfo {volume} {43}},\ \bibinfo {pages} {123001} (\bibinfo
  {year} {2026})},\ \Eprint {http://arxiv.org/abs/2505.23895} {arXiv:2505.23895
  [gr-qc]} \BibitemShut {NoStop}%
\bibitem [{\citenamefont {Franchini}\ and\ \citenamefont
  {V{\"o}lkel}(2024)}]{Franchini:2023eda}%
  \BibitemOpen
  \bibfield  {author} {\bibinfo {author} {\bibfnamefont {N.}~\bibnamefont
  {Franchini}} and \bibinfo {author} {\bibfnamefont {S.~H.}\ \bibnamefont
  {V{\"o}lkel}},\ }\enquote {\bibinfo {title} {{Testing General Relativity with
  Black Hole Quasi-normal Modes}},}\ \ (\bibinfo {year} {2024})\ \Eprint
  {http://arxiv.org/abs/2305.01696} {arXiv:2305.01696 [gr-qc]} \BibitemShut
  {NoStop}%
\bibitem [{\citenamefont
  {Vishveshwara}(1970{\natexlab{b}})}]{Vishveshwara:1970cc}%
  \BibitemOpen
  \bibfield  {author} {\bibinfo {author} {\bibfnamefont {C.~V.}\ \bibnamefont
  {Vishveshwara}},\ }\href {\doibase 10.1103/PhysRevD.1.2870} {\bibfield
  {journal} {\bibinfo  {journal} {\emph {Phys. Rev. D}}\ }\textbf {\bibinfo
  {volume} {1}},\ \bibinfo {pages} {2870} (\bibinfo {year}
  {1970}{\natexlab{b}})}\BibitemShut {NoStop}%
\bibitem [{\citenamefont {Press}(1971)}]{Press:1971wr}%
  \BibitemOpen
  \bibfield  {author} {\bibinfo {author} {\bibfnamefont {W.~H.}\ \bibnamefont
  {Press}},\ }\href {\doibase 10.1086/180849} {\bibfield  {journal} {\bibinfo
  {journal} {\emph {Astrophys. J. Lett.}}\ }\textbf {\bibinfo {volume} {170}},\
  \bibinfo {pages} {L105} (\bibinfo {year} {1971})}\BibitemShut {NoStop}%
\bibitem [{\citenamefont
  {Chandrasekhar}(1983)}]{chandrasekhar1983mathematical}%
  \BibitemOpen
  \bibfield  {author} {\bibinfo {author} {\bibfnamefont {S.}~\bibnamefont
  {Chandrasekhar}},\ }\href@noop {} {\emph {\bibinfo {title} {The Mathematical
  Theory of Black Holes}}},\ International Series of Monographs on Physics\
  (\bibinfo  {publisher} {Clarendon Press},\ \bibinfo {address} {Oxford},\
  \bibinfo {year} {1983})\BibitemShut {NoStop}%
\bibitem [{\citenamefont {Amaro-Seoane}\ \emph {et~al.}(2017)\citenamefont
  {Amaro-Seoane}, \citenamefont {Audley}, \citenamefont {Babak}, \citenamefont
  {Baker}, \citenamefont {Barausse}, \citenamefont {Bender}, \citenamefont
  {Berti}, \citenamefont {Binetruy}, \citenamefont {Born}, \citenamefont
  {Bortoluzzi} \emph {et~al.}}]{amaro2017laser}%
  \BibitemOpen
  \bibfield  {author} {\bibinfo {author} {\bibfnamefont {P.}~\bibnamefont
  {Amaro-Seoane}}, \bibinfo {author} {\bibfnamefont {H.}~\bibnamefont
  {Audley}}, \bibinfo {author} {\bibfnamefont {S.}~\bibnamefont {Babak}},
  \bibinfo {author} {\bibfnamefont {J.}~\bibnamefont {Baker}}, \bibinfo
  {author} {\bibfnamefont {E.}~\bibnamefont {Barausse}}, \bibinfo {author}
  {\bibfnamefont {P.}~\bibnamefont {Bender}}, \bibinfo {author} {\bibfnamefont
  {E.}~\bibnamefont {Berti}}, \bibinfo {author} {\bibfnamefont
  {P.}~\bibnamefont {Binetruy}}, \bibinfo {author} {\bibfnamefont
  {M.}~\bibnamefont {Born}}, \bibinfo {author} {\bibfnamefont {D.}~\bibnamefont
  {Bortoluzzi}},  \emph {et~al.},\ }\href@noop {} {\bibfield  {journal}
  {\bibinfo  {journal} {\emph {arXiv preprint arXiv:1702.00786}}\ } (\bibinfo
  {year} {2017})}\BibitemShut {NoStop}%
\bibitem [{\citenamefont {Baibhav}\ \emph {et~al.}(2020)\citenamefont
  {Baibhav}, \citenamefont {Berti},\ and\ \citenamefont
  {Cardoso}}]{Baibhav:2020tma}%
  \BibitemOpen
  \bibfield  {author} {\bibinfo {author} {\bibfnamefont {V.}~\bibnamefont
  {Baibhav}}, \bibinfo {author} {\bibfnamefont {E.}~\bibnamefont {Berti}},  and
  \bibinfo {author} {\bibfnamefont {V.}~\bibnamefont {Cardoso}},\ }\href
  {\doibase 10.1103/PhysRevD.101.084053} {\bibfield  {journal} {\bibinfo
  {journal} {\emph {Phys. Rev. D}}\ }\textbf {\bibinfo {volume} {101}},\
  \bibinfo {pages} {084053} (\bibinfo {year} {2020})},\ \Eprint
  {http://arxiv.org/abs/2001.10011} {arXiv:2001.10011 [gr-qc]} \BibitemShut
  {NoStop}%
\bibitem [{\citenamefont {Piro}\ \emph {et~al.}(2023)\citenamefont {Piro} \emph
  {et~al.}}]{Piro:2022zos}%
  \BibitemOpen
  \bibfield  {author} {\bibinfo {author} {\bibfnamefont {L.}~\bibnamefont
  {Piro}} \emph {et~al.},\ }\href {\doibase 10.1093/mnras/stad659} {\bibfield
  {journal} {\bibinfo  {journal} {\emph {Mon. Not. Roy. Astron. Soc.}}\
  }\textbf {\bibinfo {volume} {521}},\ \bibinfo {pages} {2577} (\bibinfo {year}
  {2023})},\ \Eprint {http://arxiv.org/abs/2211.13759} {arXiv:2211.13759
  [astro-ph.HE]} \BibitemShut {NoStop}%
\bibitem [{\citenamefont {Seoane}\ \emph {et~al.}(2023)\citenamefont {Seoane}
  \emph {et~al.}}]{LISA:2022yao}%
  \BibitemOpen
  \bibfield  {author} {\bibinfo {author} {\bibfnamefont {P.~A.}\ \bibnamefont
  {Seoane}} \emph {et~al.} (\bibinfo {collaboration} {LISA}),\ }\href {\doibase
  10.1007/s41114-022-00041-y} {\bibfield  {journal} {\bibinfo  {journal} {\emph
  {Living Rev. Rel.}}\ }\textbf {\bibinfo {volume} {26}},\ \bibinfo {pages} {2}
  (\bibinfo {year} {2023})},\ \Eprint {http://arxiv.org/abs/2203.06016}
  {arXiv:2203.06016 [gr-qc]} \BibitemShut {NoStop}%
\bibitem [{\citenamefont {Deng}\ \emph {et~al.}(2025)\citenamefont {Deng},
  \citenamefont {Babak},\ and\ \citenamefont {Marsat}}]{Deng:2025qhx}%
  \BibitemOpen
  \bibfield  {author} {\bibinfo {author} {\bibfnamefont {S.}~\bibnamefont
  {Deng}}, \bibinfo {author} {\bibfnamefont {S.}~\bibnamefont {Babak}},  and
  \bibinfo {author} {\bibfnamefont {S.}~\bibnamefont {Marsat}},\ }\href
  {\doibase 10.1103/jyr7-fcgp} {\bibfield  {journal} {\bibinfo  {journal}
  {\emph {Phys. Rev. D}}\ }\textbf {\bibinfo {volume} {112}},\ \bibinfo {pages}
  {043010} (\bibinfo {year} {2025})},\ \Eprint
  {http://arxiv.org/abs/2504.11322} {arXiv:2504.11322 [gr-qc]} \BibitemShut
  {NoStop}%
\bibitem [{\citenamefont {Freese}(2009)}]{Freese:2008cz}%
  \BibitemOpen
  \bibfield  {author} {\bibinfo {author} {\bibfnamefont {K.}~\bibnamefont
  {Freese}},\ }\href {\doibase 10.1051/eas/0936016} {\bibfield  {journal}
  {\bibinfo  {journal} {\emph {EAS Publ. Ser.}}\ }\textbf {\bibinfo {volume}
  {36}},\ \bibinfo {pages} {113} (\bibinfo {year} {2009})},\ \Eprint
  {http://arxiv.org/abs/0812.4005} {arXiv:0812.4005 [astro-ph]} \BibitemShut
  {NoStop}%
\bibitem [{\citenamefont {Navarro}\ \emph {et~al.}(1996)\citenamefont
  {Navarro}, \citenamefont {Frenk},\ and\ \citenamefont
  {White}}]{Navarro:1995iw}%
  \BibitemOpen
  \bibfield  {author} {\bibinfo {author} {\bibfnamefont {J.~F.}\ \bibnamefont
  {Navarro}}, \bibinfo {author} {\bibfnamefont {C.~S.}\ \bibnamefont {Frenk}},
  and \bibinfo {author} {\bibfnamefont {S.~D.~M.}\ \bibnamefont {White}},\
  }\href {\doibase 10.1086/177173} {\bibfield  {journal} {\bibinfo  {journal}
  {\emph {Astrophys. J.}}\ }\textbf {\bibinfo {volume} {462}},\ \bibinfo
  {pages} {563} (\bibinfo {year} {1996})},\ \Eprint
  {http://arxiv.org/abs/astro-ph/9508025} {arXiv:astro-ph/9508025} \BibitemShut
  {NoStop}%
\bibitem [{\citenamefont {Clowe}\ \emph {et~al.}(2006)\citenamefont {Clowe},
  \citenamefont {Bradac}, \citenamefont {Gonzalez}, \citenamefont {Markevitch},
  \citenamefont {Randall}, \citenamefont {Jones},\ and\ \citenamefont
  {Zaritsky}}]{Clowe:2006eq}%
  \BibitemOpen
  \bibfield  {author} {\bibinfo {author} {\bibfnamefont {D.}~\bibnamefont
  {Clowe}}, \bibinfo {author} {\bibfnamefont {M.}~\bibnamefont {Bradac}},
  \bibinfo {author} {\bibfnamefont {A.~H.}\ \bibnamefont {Gonzalez}}, \bibinfo
  {author} {\bibfnamefont {M.}~\bibnamefont {Markevitch}}, \bibinfo {author}
  {\bibfnamefont {S.~W.}\ \bibnamefont {Randall}}, \bibinfo {author}
  {\bibfnamefont {C.}~\bibnamefont {Jones}},  and \bibinfo {author}
  {\bibfnamefont {D.}~\bibnamefont {Zaritsky}},\ }\href {\doibase
  10.1086/508162} {\bibfield  {journal} {\bibinfo  {journal} {\emph {Astrophys.
  J. Lett.}}\ }\textbf {\bibinfo {volume} {648}},\ \bibinfo {pages} {L109}
  (\bibinfo {year} {2006})},\ \Eprint {http://arxiv.org/abs/astro-ph/0608407}
  {arXiv:astro-ph/0608407} \BibitemShut {NoStop}%
\bibitem [{\citenamefont {Bertone}\ \emph {et~al.}(2005)\citenamefont
  {Bertone}, \citenamefont {Hooper},\ and\ \citenamefont
  {Silk}}]{Bertone:2004pz}%
  \BibitemOpen
  \bibfield  {author} {\bibinfo {author} {\bibfnamefont {G.}~\bibnamefont
  {Bertone}}, \bibinfo {author} {\bibfnamefont {D.}~\bibnamefont {Hooper}},
  and \bibinfo {author} {\bibfnamefont {J.}~\bibnamefont {Silk}},\ }\href
  {\doibase 10.1016/j.physrep.2004.08.031} {\bibfield  {journal} {\bibinfo
  {journal} {\emph {Phys. Rept.}}\ }\textbf {\bibinfo {volume} {405}},\
  \bibinfo {pages} {279} (\bibinfo {year} {2005})},\ \Eprint
  {http://arxiv.org/abs/hep-ph/0404175} {arXiv:hep-ph/0404175} \BibitemShut
  {NoStop}%
\bibitem [{\citenamefont {Kahlhoefer}(2017)}]{Kahlhoefer:2017dnp}%
  \BibitemOpen
  \bibfield  {author} {\bibinfo {author} {\bibfnamefont {F.}~\bibnamefont
  {Kahlhoefer}},\ }\href {\doibase 10.1142/S0217751X1730006X} {\bibfield
  {journal} {\bibinfo  {journal} {\emph {Int. J. Mod. Phys. A}}\ }\textbf
  {\bibinfo {volume} {32}},\ \bibinfo {pages} {1730006} (\bibinfo {year}
  {2017})},\ \Eprint {http://arxiv.org/abs/1702.02430} {arXiv:1702.02430
  [hep-ph]} \BibitemShut {NoStop}%
\bibitem [{\citenamefont {P{\'e}rez de~los
  Heros}(2020)}]{PerezdelosHeros:2020qyt}%
  \BibitemOpen
  \bibfield  {author} {\bibinfo {author} {\bibfnamefont {C.}~\bibnamefont
  {P{\'e}rez de~los Heros}},\ }\href {\doibase 10.3390/sym12101648} {\bibfield
  {journal} {\bibinfo  {journal} {\emph {Symmetry}}\ }\textbf {\bibinfo
  {volume} {12}},\ \bibinfo {pages} {1648} (\bibinfo {year} {2020})},\ \Eprint
  {http://arxiv.org/abs/2008.11561} {arXiv:2008.11561 [astro-ph.HE]}
  \BibitemShut {NoStop}%
\bibitem [{\citenamefont {Ghez}\ \emph {et~al.}(2008)\citenamefont {Ghez} \emph
  {et~al.}}]{Ghez:2008ms}%
  \BibitemOpen
  \bibfield  {author} {\bibinfo {author} {\bibfnamefont {A.~M.}\ \bibnamefont
  {Ghez}} \emph {et~al.},\ }\href {\doibase 10.1086/592738} {\bibfield
  {journal} {\bibinfo  {journal} {\emph {Astrophys. J.}}\ }\textbf {\bibinfo
  {volume} {689}},\ \bibinfo {pages} {1044} (\bibinfo {year} {2008})},\ \Eprint
  {http://arxiv.org/abs/0808.2870} {arXiv:0808.2870 [astro-ph]} \BibitemShut
  {NoStop}%
\bibitem [{\citenamefont {Gillessen}\ \emph {et~al.}(2009)\citenamefont
  {Gillessen}, \citenamefont {Eisenhauer}, \citenamefont {Trippe},
  \citenamefont {Alexander}, \citenamefont {Genzel}, \citenamefont {Martins},\
  and\ \citenamefont {Ott}}]{Gillessen:2008qv}%
  \BibitemOpen
  \bibfield  {author} {\bibinfo {author} {\bibfnamefont {S.}~\bibnamefont
  {Gillessen}}, \bibinfo {author} {\bibfnamefont {F.}~\bibnamefont
  {Eisenhauer}}, \bibinfo {author} {\bibfnamefont {S.}~\bibnamefont {Trippe}},
  \bibinfo {author} {\bibfnamefont {T.}~\bibnamefont {Alexander}}, \bibinfo
  {author} {\bibfnamefont {R.}~\bibnamefont {Genzel}}, \bibinfo {author}
  {\bibfnamefont {F.}~\bibnamefont {Martins}},  and \bibinfo {author}
  {\bibfnamefont {T.}~\bibnamefont {Ott}},\ }\href {\doibase
  10.1088/0004-637X/692/2/1075} {\bibfield  {journal} {\bibinfo  {journal}
  {\emph {Astrophys. J.}}\ }\textbf {\bibinfo {volume} {692}},\ \bibinfo
  {pages} {1075} (\bibinfo {year} {2009})},\ \Eprint
  {http://arxiv.org/abs/0810.4674} {arXiv:0810.4674 [astro-ph]} \BibitemShut
  {NoStop}%
\bibitem [{\citenamefont {Miyoshi}\ \emph {et~al.}(1995)\citenamefont
  {Miyoshi}, \citenamefont {Moran}, \citenamefont {Hernstein}, \citenamefont
  {Greenhill}, \citenamefont {Nakai}, \citenamefont {Diamond},\ and\
  \citenamefont {Inoue}}]{Miyoshi:1995da}%
  \BibitemOpen
  \bibfield  {author} {\bibinfo {author} {\bibfnamefont {m.}~\bibnamefont
  {Miyoshi}}, \bibinfo {author} {\bibfnamefont {J.}~\bibnamefont {Moran}},
  \bibinfo {author} {\bibfnamefont {J.}~\bibnamefont {Hernstein}}, \bibinfo
  {author} {\bibfnamefont {L.}~\bibnamefont {Greenhill}}, \bibinfo {author}
  {\bibfnamefont {N.}~\bibnamefont {Nakai}}, \bibinfo {author} {\bibfnamefont
  {P.}~\bibnamefont {Diamond}},  and \bibinfo {author} {\bibfnamefont
  {M.}~\bibnamefont {Inoue}},\ }\href {\doibase 10.1038/373127a0} {\bibfield
  {journal} {\bibinfo  {journal} {\emph {Nature}}\ }\textbf {\bibinfo {volume}
  {373}},\ \bibinfo {pages} {127} (\bibinfo {year} {1995})}\BibitemShut
  {NoStop}%
\bibitem [{\citenamefont {Kormendy}\ and\ \citenamefont
  {Ho}(2013)}]{Kormendy:2013dxa}%
  \BibitemOpen
  \bibfield  {author} {\bibinfo {author} {\bibfnamefont {J.}~\bibnamefont
  {Kormendy}} and \bibinfo {author} {\bibfnamefont {L.~C.}\ \bibnamefont
  {Ho}},\ }\href {\doibase 10.1146/annurev-astro-082708-101811} {\bibfield
  {journal} {\bibinfo  {journal} {\emph {Ann. Rev. Astron. Astrophys.}}\
  }\textbf {\bibinfo {volume} {51}},\ \bibinfo {pages} {511} (\bibinfo {year}
  {2013})},\ \Eprint {http://arxiv.org/abs/1304.7762} {arXiv:1304.7762
  [astro-ph.CO]} \BibitemShut {NoStop}%
\bibitem [{\citenamefont {Ferrarese}\ and\ \citenamefont
  {Merritt}(2000)}]{Ferrarese:2000se}%
  \BibitemOpen
  \bibfield  {author} {\bibinfo {author} {\bibfnamefont {L.}~\bibnamefont
  {Ferrarese}} and \bibinfo {author} {\bibfnamefont {D.}~\bibnamefont
  {Merritt}},\ }\href {\doibase 10.1086/312838} {\bibfield  {journal} {\bibinfo
   {journal} {\emph {Astrophys. J. Lett.}}\ }\textbf {\bibinfo {volume}
  {539}},\ \bibinfo {pages} {L9} (\bibinfo {year} {2000})},\ \Eprint
  {http://arxiv.org/abs/astro-ph/0006053} {arXiv:astro-ph/0006053} \BibitemShut
  {NoStop}%
\bibitem [{\citenamefont {Gebhardt}\ \emph {et~al.}(2000)\citenamefont
  {Gebhardt} \emph {et~al.}}]{Gebhardt:2000fk}%
  \BibitemOpen
  \bibfield  {author} {\bibinfo {author} {\bibfnamefont {K.}~\bibnamefont
  {Gebhardt}} \emph {et~al.},\ }\href {\doibase 10.1086/312840} {\bibfield
  {journal} {\bibinfo  {journal} {\emph {Astrophys. J. Lett.}}\ }\textbf
  {\bibinfo {volume} {539}},\ \bibinfo {pages} {L13} (\bibinfo {year}
  {2000})},\ \Eprint {http://arxiv.org/abs/astro-ph/0006289}
  {arXiv:astro-ph/0006289} \BibitemShut {NoStop}%
\bibitem [{\citenamefont {Navarro}\ \emph {et~al.}(1997)\citenamefont
  {Navarro}, \citenamefont {Frenk},\ and\ \citenamefont
  {White}}]{Navarro:1996gj}%
  \BibitemOpen
  \bibfield  {author} {\bibinfo {author} {\bibfnamefont {J.~F.}\ \bibnamefont
  {Navarro}}, \bibinfo {author} {\bibfnamefont {C.~S.}\ \bibnamefont {Frenk}},
  and \bibinfo {author} {\bibfnamefont {S.~D.~M.}\ \bibnamefont {White}},\
  }\href {\doibase 10.1086/304888} {\bibfield  {journal} {\bibinfo  {journal}
  {\emph {Astrophys. J.}}\ }\textbf {\bibinfo {volume} {490}},\ \bibinfo
  {pages} {493} (\bibinfo {year} {1997})},\ \Eprint
  {http://arxiv.org/abs/astro-ph/9611107} {arXiv:astro-ph/9611107} \BibitemShut
  {NoStop}%
\bibitem [{\citenamefont {Salucci}(2019)}]{Salucci:2018hqu}%
  \BibitemOpen
  \bibfield  {author} {\bibinfo {author} {\bibfnamefont {P.}~\bibnamefont
  {Salucci}},\ }\href {\doibase 10.1007/s00159-018-0113-1} {\bibfield
  {journal} {\bibinfo  {journal} {\emph {Astron. Astrophys. Rev.}}\ }\textbf
  {\bibinfo {volume} {27}},\ \bibinfo {pages} {2} (\bibinfo {year} {2019})},\
  \Eprint {http://arxiv.org/abs/1811.08843} {arXiv:1811.08843 [astro-ph.GA]}
  \BibitemShut {NoStop}%
\bibitem [{\citenamefont {Gondolo}\ and\ \citenamefont
  {Silk}(1999)}]{Gondolo:1999ef}%
  \BibitemOpen
  \bibfield  {author} {\bibinfo {author} {\bibfnamefont {P.}~\bibnamefont
  {Gondolo}} and \bibinfo {author} {\bibfnamefont {J.}~\bibnamefont {Silk}},\
  }\href {\doibase 10.1103/PhysRevLett.83.1719} {\bibfield  {journal} {\bibinfo
   {journal} {\emph {Phys. Rev. Lett.}}\ }\textbf {\bibinfo {volume} {83}},\
  \bibinfo {pages} {1719} (\bibinfo {year} {1999})},\ \Eprint
  {http://arxiv.org/abs/astro-ph/9906391} {arXiv:astro-ph/9906391} \BibitemShut
  {NoStop}%
\bibitem [{\citenamefont {Sadeghian}\ \emph {et~al.}(2013)\citenamefont
  {Sadeghian}, \citenamefont {Ferrer},\ and\ \citenamefont
  {Will}}]{Sadeghian:2013laa}%
  \BibitemOpen
  \bibfield  {author} {\bibinfo {author} {\bibfnamefont {L.}~\bibnamefont
  {Sadeghian}}, \bibinfo {author} {\bibfnamefont {F.}~\bibnamefont {Ferrer}},
  and \bibinfo {author} {\bibfnamefont {C.~M.}\ \bibnamefont {Will}},\ }\href
  {\doibase 10.1103/PhysRevD.88.063522} {\bibfield  {journal} {\bibinfo
  {journal} {\emph {Phys. Rev. D}}\ }\textbf {\bibinfo {volume} {88}},\
  \bibinfo {pages} {063522} (\bibinfo {year} {2013})},\ \Eprint
  {http://arxiv.org/abs/1305.2619} {arXiv:1305.2619 [astro-ph.GA]} \BibitemShut
  {NoStop}%
\bibitem [{\citenamefont {Merritt}\ \emph {et~al.}(2002)\citenamefont
  {Merritt}, \citenamefont {Milosavljevic}, \citenamefont {Verde},\ and\
  \citenamefont {Jimenez}}]{Merritt:2002vj}%
  \BibitemOpen
  \bibfield  {author} {\bibinfo {author} {\bibfnamefont {D.}~\bibnamefont
  {Merritt}}, \bibinfo {author} {\bibfnamefont {M.}~\bibnamefont
  {Milosavljevic}}, \bibinfo {author} {\bibfnamefont {L.}~\bibnamefont
  {Verde}},  and \bibinfo {author} {\bibfnamefont {R.}~\bibnamefont
  {Jimenez}},\ }\href {\doibase 10.1103/PhysRevLett.88.191301} {\bibfield
  {journal} {\bibinfo  {journal} {\emph {Phys. Rev. Lett.}}\ }\textbf {\bibinfo
  {volume} {88}},\ \bibinfo {pages} {191301} (\bibinfo {year} {2002})},\
  \Eprint {http://arxiv.org/abs/astro-ph/0201376} {arXiv:astro-ph/0201376}
  \BibitemShut {NoStop}%
\bibitem [{\citenamefont {Merritt}(2004)}]{Merritt:2003qk}%
  \BibitemOpen
  \bibfield  {author} {\bibinfo {author} {\bibfnamefont {D.}~\bibnamefont
  {Merritt}},\ }\href {\doibase 10.1103/PhysRevLett.92.201304} {\bibfield
  {journal} {\bibinfo  {journal} {\emph {Phys. Rev. Lett.}}\ }\textbf {\bibinfo
  {volume} {92}},\ \bibinfo {pages} {201304} (\bibinfo {year} {2004})},\
  \Eprint {http://arxiv.org/abs/astro-ph/0311594} {arXiv:astro-ph/0311594}
  \BibitemShut {NoStop}%
\bibitem [{\citenamefont {Ullio}\ \emph {et~al.}(2001)\citenamefont {Ullio},
  \citenamefont {Zhao},\ and\ \citenamefont {Kamionkowski}}]{Ullio:2001fb}%
  \BibitemOpen
  \bibfield  {author} {\bibinfo {author} {\bibfnamefont {P.}~\bibnamefont
  {Ullio}}, \bibinfo {author} {\bibfnamefont {H.}~\bibnamefont {Zhao}},  and
  \bibinfo {author} {\bibfnamefont {M.}~\bibnamefont {Kamionkowski}},\ }\href
  {\doibase 10.1103/PhysRevD.64.043504} {\bibfield  {journal} {\bibinfo
  {journal} {\emph {Phys. Rev. D}}\ }\textbf {\bibinfo {volume} {64}},\
  \bibinfo {pages} {043504} (\bibinfo {year} {2001})},\ \Eprint
  {http://arxiv.org/abs/astro-ph/0101481} {arXiv:astro-ph/0101481} \BibitemShut
  {NoStop}%
\bibitem [{\citenamefont {Bertone}\ and\ \citenamefont
  {Merritt}(2005)}]{Bertone:2005hw}%
  \BibitemOpen
  \bibfield  {author} {\bibinfo {author} {\bibfnamefont {G.}~\bibnamefont
  {Bertone}} and \bibinfo {author} {\bibfnamefont {D.}~\bibnamefont
  {Merritt}},\ }\href {\doibase 10.1103/PhysRevD.72.103502} {\bibfield
  {journal} {\bibinfo  {journal} {\emph {Phys. Rev. D}}\ }\textbf {\bibinfo
  {volume} {72}},\ \bibinfo {pages} {103502} (\bibinfo {year} {2005})},\
  \Eprint {http://arxiv.org/abs/astro-ph/0501555} {arXiv:astro-ph/0501555}
  \BibitemShut {NoStop}%
\bibitem [{\citenamefont {Barack}\ \emph {et~al.}(2019)\citenamefont {Barack}
  \emph {et~al.}}]{Barack:2018yly}%
  \BibitemOpen
  \bibfield  {author} {\bibinfo {author} {\bibfnamefont {L.}~\bibnamefont
  {Barack}} \emph {et~al.},\ }\href {\doibase 10.1088/1361-6382/ab0587}
  {\bibfield  {journal} {\bibinfo  {journal} {\emph {Class. Quant. Grav.}}\
  }\textbf {\bibinfo {volume} {36}},\ \bibinfo {pages} {143001} (\bibinfo
  {year} {2019})},\ \Eprint {http://arxiv.org/abs/1806.05195} {arXiv:1806.05195
  [gr-qc]} \BibitemShut {NoStop}%
\bibitem [{\citenamefont {Eda}\ \emph {et~al.}(2013)\citenamefont {Eda},
  \citenamefont {Itoh}, \citenamefont {Kuroyanagi},\ and\ \citenamefont
  {Silk}}]{Eda:2013gg}%
  \BibitemOpen
  \bibfield  {author} {\bibinfo {author} {\bibfnamefont {K.}~\bibnamefont
  {Eda}}, \bibinfo {author} {\bibfnamefont {Y.}~\bibnamefont {Itoh}}, \bibinfo
  {author} {\bibfnamefont {S.}~\bibnamefont {Kuroyanagi}},  and \bibinfo
  {author} {\bibfnamefont {J.}~\bibnamefont {Silk}},\ }\href {\doibase
  10.1103/PhysRevLett.110.221101} {\bibfield  {journal} {\bibinfo  {journal}
  {\emph {Phys. Rev. Lett.}}\ }\textbf {\bibinfo {volume} {110}},\ \bibinfo
  {pages} {221101} (\bibinfo {year} {2013})},\ \Eprint
  {http://arxiv.org/abs/1301.5971} {arXiv:1301.5971 [gr-qc]} \BibitemShut
  {NoStop}%
\bibitem [{\citenamefont {Macedo}\ \emph {et~al.}(2013)\citenamefont {Macedo},
  \citenamefont {Pani}, \citenamefont {Cardoso},\ and\ \citenamefont
  {Crispino}}]{Macedo:2013qea}%
  \BibitemOpen
  \bibfield  {author} {\bibinfo {author} {\bibfnamefont {C.~F.~B.}\
  \bibnamefont {Macedo}}, \bibinfo {author} {\bibfnamefont {P.}~\bibnamefont
  {Pani}}, \bibinfo {author} {\bibfnamefont {V.}~\bibnamefont {Cardoso}},  and
  \bibinfo {author} {\bibfnamefont {L.~C.~B.}\ \bibnamefont {Crispino}},\
  }\href {\doibase 10.1088/0004-637X/774/1/48} {\bibfield  {journal} {\bibinfo
  {journal} {\emph {Astrophys. J.}}\ }\textbf {\bibinfo {volume} {774}},\
  \bibinfo {pages} {48} (\bibinfo {year} {2013})},\ \Eprint
  {http://arxiv.org/abs/1302.2646} {arXiv:1302.2646 [gr-qc]} \BibitemShut
  {NoStop}%
\bibitem [{\citenamefont {Barausse}\ \emph {et~al.}(2014)\citenamefont
  {Barausse}, \citenamefont {Cardoso},\ and\ \citenamefont
  {Pani}}]{Barausse:2014tra}%
  \BibitemOpen
  \bibfield  {author} {\bibinfo {author} {\bibfnamefont {E.}~\bibnamefont
  {Barausse}}, \bibinfo {author} {\bibfnamefont {V.}~\bibnamefont {Cardoso}},
  and \bibinfo {author} {\bibfnamefont {P.}~\bibnamefont {Pani}},\ }\href
  {\doibase 10.1103/PhysRevD.89.104059} {\bibfield  {journal} {\bibinfo
  {journal} {\emph {Phys. Rev. D}}\ }\textbf {\bibinfo {volume} {89}},\
  \bibinfo {pages} {104059} (\bibinfo {year} {2014})},\ \Eprint
  {http://arxiv.org/abs/1404.7149} {arXiv:1404.7149 [gr-qc]} \BibitemShut
  {NoStop}%
\bibitem [{\citenamefont {Baibhav}\ \emph {et~al.}(2021)\citenamefont {Baibhav}
  \emph {et~al.}}]{Baibhav:2019rsa}%
  \BibitemOpen
  \bibfield  {author} {\bibinfo {author} {\bibfnamefont {V.}~\bibnamefont
  {Baibhav}} \emph {et~al.},\ }\href {\doibase 10.1007/s10686-021-09741-9}
  {\bibfield  {journal} {\bibinfo  {journal} {\emph {Exper. Astron.}}\ }\textbf
  {\bibinfo {volume} {51}},\ \bibinfo {pages} {1385} (\bibinfo {year}
  {2021})},\ \Eprint {http://arxiv.org/abs/1908.11390} {arXiv:1908.11390
  [astro-ph.HE]} \BibitemShut {NoStop}%
\bibitem [{\citenamefont {Seoane}\ \emph {et~al.}(2022)\citenamefont {Seoane}
  \emph {et~al.}}]{Seoane:2021kkk}%
  \BibitemOpen
  \bibfield  {author} {\bibinfo {author} {\bibfnamefont {P.~A.}\ \bibnamefont
  {Seoane}} \emph {et~al.},\ }\href {\doibase 10.1007/s10714-021-02889-x}
  {\bibfield  {journal} {\bibinfo  {journal} {\emph {Gen. Rel. Grav.}}\
  }\textbf {\bibinfo {volume} {54}},\ \bibinfo {pages} {3} (\bibinfo {year}
  {2022})},\ \Eprint {http://arxiv.org/abs/2107.09665} {arXiv:2107.09665
  [astro-ph.IM]} \BibitemShut {NoStop}%
\bibitem [{\citenamefont {Cardoso}\ and\ \citenamefont
  {Maselli}(2020)}]{Cardoso:2019rou}%
  \BibitemOpen
  \bibfield  {author} {\bibinfo {author} {\bibfnamefont {V.}~\bibnamefont
  {Cardoso}} and \bibinfo {author} {\bibfnamefont {A.}~\bibnamefont
  {Maselli}},\ }\href {\doibase 10.1051/0004-6361/202037654} {\bibfield
  {journal} {\bibinfo  {journal} {\emph {Astron. Astrophys.}}\ }\textbf
  {\bibinfo {volume} {644}},\ \bibinfo {pages} {A147} (\bibinfo {year}
  {2020})},\ \Eprint {http://arxiv.org/abs/1909.05870} {arXiv:1909.05870
  [astro-ph.HE]} \BibitemShut {NoStop}%
\bibitem [{\citenamefont {Kavanagh}\ \emph {et~al.}(2020)\citenamefont
  {Kavanagh}, \citenamefont {Nichols}, \citenamefont {Bertone},\ and\
  \citenamefont {Gaggero}}]{Kavanagh:2020cfn}%
  \BibitemOpen
  \bibfield  {author} {\bibinfo {author} {\bibfnamefont {B.~J.}\ \bibnamefont
  {Kavanagh}}, \bibinfo {author} {\bibfnamefont {D.~A.}\ \bibnamefont
  {Nichols}}, \bibinfo {author} {\bibfnamefont {G.}~\bibnamefont {Bertone}},
  and \bibinfo {author} {\bibfnamefont {D.}~\bibnamefont {Gaggero}},\ }\href
  {\doibase 10.1103/PhysRevD.102.083006} {\bibfield  {journal} {\bibinfo
  {journal} {\emph {Phys. Rev. D}}\ }\textbf {\bibinfo {volume} {102}},\
  \bibinfo {pages} {083006} (\bibinfo {year} {2020})},\ \Eprint
  {http://arxiv.org/abs/2002.12811} {arXiv:2002.12811 [gr-qc]} \BibitemShut
  {NoStop}%
\bibitem [{\citenamefont {Tamanini}\ \emph {et~al.}(2020)\citenamefont
  {Tamanini}, \citenamefont {Klein}, \citenamefont {Bonvin}, \citenamefont
  {Barausse},\ and\ \citenamefont {Caprini}}]{Tamanini:2019usx}%
  \BibitemOpen
  \bibfield  {author} {\bibinfo {author} {\bibfnamefont {N.}~\bibnamefont
  {Tamanini}}, \bibinfo {author} {\bibfnamefont {A.}~\bibnamefont {Klein}},
  \bibinfo {author} {\bibfnamefont {C.}~\bibnamefont {Bonvin}}, \bibinfo
  {author} {\bibfnamefont {E.}~\bibnamefont {Barausse}},  and \bibinfo {author}
  {\bibfnamefont {C.}~\bibnamefont {Caprini}},\ }\href {\doibase
  10.1103/PhysRevD.101.063002} {\bibfield  {journal} {\bibinfo  {journal}
  {\emph {Phys. Rev. D}}\ }\textbf {\bibinfo {volume} {101}},\ \bibinfo {pages}
  {063002} (\bibinfo {year} {2020})},\ \Eprint
  {http://arxiv.org/abs/1907.02018} {arXiv:1907.02018 [astro-ph.IM]}
  \BibitemShut {NoStop}%
\bibitem [{\citenamefont {Chowdhury}\ \emph {et~al.}(2025)\citenamefont
  {Chowdhury}, \citenamefont {Sen}, \citenamefont {Chakrabarti},\ and\
  \citenamefont {Das}}]{Chowdhury:2025tpt}%
  \BibitemOpen
  \bibfield  {author} {\bibinfo {author} {\bibfnamefont {A.}~\bibnamefont
  {Chowdhury}}, \bibinfo {author} {\bibfnamefont {G.}~\bibnamefont {Sen}},
  \bibinfo {author} {\bibfnamefont {S.}~\bibnamefont {Chakrabarti}},  and
  \bibinfo {author} {\bibfnamefont {S.}~\bibnamefont {Das}},\ }\href {\doibase
  10.1103/vqjm-3dt8} {\bibfield  {journal} {\bibinfo  {journal} {\emph {Phys.
  Rev. D}}\ }\textbf {\bibinfo {volume} {112}},\ \bibinfo {pages} {064041}
  (\bibinfo {year} {2025})},\ \Eprint {http://arxiv.org/abs/2503.08528}
  {arXiv:2503.08528 [gr-qc]} \BibitemShut {NoStop}%
\bibitem [{\citenamefont {Dosopoulou}\ and\ \citenamefont
  {Silk}(2025)}]{Dosopoulou:2025jth}%
  \BibitemOpen
  \bibfield  {author} {\bibinfo {author} {\bibfnamefont {F.}~\bibnamefont
  {Dosopoulou}} and \bibinfo {author} {\bibfnamefont {J.}~\bibnamefont
  {Silk}},\ }\href {\doibase 10.1103/hysg-5273} {\bibfield  {journal} {\bibinfo
   {journal} {\emph {Phys. Rev. Lett.}}\ }\textbf {\bibinfo {volume} {135}},\
  \bibinfo {pages} {081401} (\bibinfo {year} {2025})},\ \Eprint
  {http://arxiv.org/abs/2502.15468} {arXiv:2502.15468 [astro-ph.HE]}
  \BibitemShut {NoStop}%
\bibitem [{\citenamefont {Hassanabadi}\ \emph {et~al.}(2026)\citenamefont
  {Hassanabadi}, \citenamefont {Chen}, \citenamefont {Zare},\ and\
  \citenamefont {Perlick}}]{Hassanabadi:2026wgc}%
  \BibitemOpen
  \bibfield  {author} {\bibinfo {author} {\bibfnamefont {H.}~\bibnamefont
  {Hassanabadi}}, \bibinfo {author} {\bibfnamefont {C.-Y.}\ \bibnamefont
  {Chen}}, \bibinfo {author} {\bibfnamefont {S.}~\bibnamefont {Zare}},  and
  \bibinfo {author} {\bibfnamefont {V.}~\bibnamefont {Perlick}},\ }\Eprint
  {http://arxiv.org/abs/2608.11430} {arXiv:2608.11430 [gr-qc]} \BibitemShut
  {NoStop}%
\bibitem [{\citenamefont {Hernquist}(1990)}]{hernquist1990analytical}%
  \BibitemOpen
  \bibfield  {author} {\bibinfo {author} {\bibfnamefont {L.}~\bibnamefont
  {Hernquist}},\ }\href@noop {} {\bibfield  {journal} {\bibinfo  {journal}
  {\emph {Astrophysical Journal, Part 1 (ISSN 0004-637X), vol. 356, June 20,
  1990, p. 359-364.}}\ }\textbf {\bibinfo {volume} {356}},\ \bibinfo {pages}
  {359} (\bibinfo {year} {1990})}\BibitemShut {NoStop}%
\bibitem [{\citenamefont {Jaffe}(1983)}]{Jaffe:1983iv}%
  \BibitemOpen
  \bibfield  {author} {\bibinfo {author} {\bibfnamefont {W.}~\bibnamefont
  {Jaffe}},\ }\href@noop {} {\bibfield  {journal} {\bibinfo  {journal} {\emph
  {Mon. Not. Roy. Astron. Soc.}}\ }\textbf {\bibinfo {volume} {202}},\ \bibinfo
  {pages} {995} (\bibinfo {year} {1983})}\BibitemShut {NoStop}%
\bibitem [{\citenamefont {King}(1962)}]{King:1962wi}%
  \BibitemOpen
  \bibfield  {author} {\bibinfo {author} {\bibfnamefont {I.}~\bibnamefont
  {King}},\ }\href {\doibase 10.1086/108756} {\bibfield  {journal} {\bibinfo
  {journal} {\emph {Astron. J.}}\ }\textbf {\bibinfo {volume} {67}},\ \bibinfo
  {pages} {471} (\bibinfo {year} {1962})}\BibitemShut {NoStop}%
\bibitem [{\citenamefont {Graham}\ \emph {et~al.}(2006)\citenamefont {Graham},
  \citenamefont {Merritt}, \citenamefont {Moore}, \citenamefont {Diemand},\
  and\ \citenamefont {Terzic}}]{Graham:2005xx}%
  \BibitemOpen
  \bibfield  {author} {\bibinfo {author} {\bibfnamefont {A.~W.}\ \bibnamefont
  {Graham}}, \bibinfo {author} {\bibfnamefont {D.}~\bibnamefont {Merritt}},
  \bibinfo {author} {\bibfnamefont {B.}~\bibnamefont {Moore}}, \bibinfo
  {author} {\bibfnamefont {J.}~\bibnamefont {Diemand}},  and \bibinfo {author}
  {\bibfnamefont {B.}~\bibnamefont {Terzic}},\ }\href {\doibase 10.1086/508988}
  {\bibfield  {journal} {\bibinfo  {journal} {\emph {Astron. J.}}\ }\textbf
  {\bibinfo {volume} {132}},\ \bibinfo {pages} {2685} (\bibinfo {year}
  {2006})},\ \Eprint {http://arxiv.org/abs/astro-ph/0509417}
  {arXiv:astro-ph/0509417} \BibitemShut {NoStop}%
\bibitem [{\citenamefont {Taylor}\ and\ \citenamefont
  {Silk}(2003)}]{Taylor:2002zd}%
  \BibitemOpen
  \bibfield  {author} {\bibinfo {author} {\bibfnamefont {J.~E.}\ \bibnamefont
  {Taylor}} and \bibinfo {author} {\bibfnamefont {J.}~\bibnamefont {Silk}},\
  }\href {\doibase 10.1046/j.1365-8711.2003.06201.x} {\bibfield  {journal}
  {\bibinfo  {journal} {\emph {Mon. Not. Roy. Astron. Soc.}}\ }\textbf
  {\bibinfo {volume} {339}},\ \bibinfo {pages} {505} (\bibinfo {year}
  {2003})},\ \Eprint {http://arxiv.org/abs/astro-ph/0207299}
  {arXiv:astro-ph/0207299} \BibitemShut {NoStop}%
\bibitem [{\citenamefont {Dekel}\ \emph {et~al.}(2017)\citenamefont {Dekel},
  \citenamefont {Ishai}, \citenamefont {Dutton},\ and\ \citenamefont
  {Maccio}}]{dekel2017dark}%
  \BibitemOpen
  \bibfield  {author} {\bibinfo {author} {\bibfnamefont {A.}~\bibnamefont
  {Dekel}}, \bibinfo {author} {\bibfnamefont {G.}~\bibnamefont {Ishai}},
  \bibinfo {author} {\bibfnamefont {A.~A.}\ \bibnamefont {Dutton}},  and
  \bibinfo {author} {\bibfnamefont {A.~V.}\ \bibnamefont {Maccio}},\
  }\href@noop {} {\bibfield  {journal} {\bibinfo  {journal} {\emph {Monthly
  Notices of the Royal Astronomical Society}}\ }\textbf {\bibinfo {volume}
  {468}},\ \bibinfo {pages} {1005} (\bibinfo {year} {2017})}\BibitemShut
  {NoStop}%
\bibitem [{\citenamefont {Zhao}(1996)}]{Zhao:1995cp}%
  \BibitemOpen
  \bibfield  {author} {\bibinfo {author} {\bibfnamefont {H.}~\bibnamefont
  {Zhao}},\ }\href {\doibase 10.1093/mnras/278.2.488} {\bibfield  {journal}
  {\bibinfo  {journal} {\emph {Mon. Not. Roy. Astron. Soc.}}\ }\textbf
  {\bibinfo {volume} {278}},\ \bibinfo {pages} {488} (\bibinfo {year}
  {1996})},\ \Eprint {http://arxiv.org/abs/astro-ph/9509122}
  {arXiv:astro-ph/9509122} \BibitemShut {NoStop}%
\bibitem [{\citenamefont {Cardoso}\ \emph {et~al.}(2022)\citenamefont
  {Cardoso}, \citenamefont {Destounis}, \citenamefont {Duque}, \citenamefont
  {Macedo},\ and\ \citenamefont {Maselli}}]{Cardoso:2021wlq}%
  \BibitemOpen
  \bibfield  {author} {\bibinfo {author} {\bibfnamefont {V.}~\bibnamefont
  {Cardoso}}, \bibinfo {author} {\bibfnamefont {K.}~\bibnamefont {Destounis}},
  \bibinfo {author} {\bibfnamefont {F.}~\bibnamefont {Duque}}, \bibinfo
  {author} {\bibfnamefont {R.~P.}\ \bibnamefont {Macedo}},  and \bibinfo
  {author} {\bibfnamefont {A.}~\bibnamefont {Maselli}},\ }\href {\doibase
  10.1103/PhysRevD.105.L061501} {\bibfield  {journal} {\bibinfo  {journal}
  {\emph {Phys. Rev. D}}\ }\textbf {\bibinfo {volume} {105}},\ \bibinfo {pages}
  {L061501} (\bibinfo {year} {2022})},\ \Eprint
  {http://arxiv.org/abs/2109.00005} {arXiv:2109.00005 [gr-qc]} \BibitemShut
  {NoStop}%
\bibitem [{\citenamefont {Maeda}\ \emph {et~al.}(2025)\citenamefont {Maeda},
  \citenamefont {Cardoso},\ and\ \citenamefont {Wang}}]{Maeda:2024tsg}%
  \BibitemOpen
  \bibfield  {author} {\bibinfo {author} {\bibfnamefont {K.-i.}\ \bibnamefont
  {Maeda}}, \bibinfo {author} {\bibfnamefont {V.}~\bibnamefont {Cardoso}},  and
  \bibinfo {author} {\bibfnamefont {A.}~\bibnamefont {Wang}},\ }\href {\doibase
  10.1103/PhysRevD.111.044060} {\bibfield  {journal} {\bibinfo  {journal}
  {\emph {Phys. Rev. D}}\ }\textbf {\bibinfo {volume} {111}},\ \bibinfo {pages}
  {044060} (\bibinfo {year} {2025})},\ \Eprint
  {http://arxiv.org/abs/2410.04175} {arXiv:2410.04175 [gr-qc]} \BibitemShut
  {NoStop}%
\bibitem [{\citenamefont {Nampalliwar}\ \emph {et~al.}(2021)\citenamefont
  {Nampalliwar}, \citenamefont {Kumar}, \citenamefont {Jusufi}, \citenamefont
  {Wu}, \citenamefont {Jamil},\ and\ \citenamefont
  {Salucci}}]{Nampalliwar:2021tyz}%
  \BibitemOpen
  \bibfield  {author} {\bibinfo {author} {\bibfnamefont {S.}~\bibnamefont
  {Nampalliwar}}, \bibinfo {author} {\bibfnamefont {S.}~\bibnamefont {Kumar}},
  \bibinfo {author} {\bibfnamefont {K.}~\bibnamefont {Jusufi}}, \bibinfo
  {author} {\bibfnamefont {Q.}~\bibnamefont {Wu}}, \bibinfo {author}
  {\bibfnamefont {M.}~\bibnamefont {Jamil}},  and \bibinfo {author}
  {\bibfnamefont {P.}~\bibnamefont {Salucci}},\ }\href {\doibase
  10.3847/1538-4357/ac05cc} {\bibfield  {journal} {\bibinfo  {journal} {\emph
  {Astrophys. J.}}\ }\textbf {\bibinfo {volume} {916}},\ \bibinfo {pages} {116}
  (\bibinfo {year} {2021})},\ \Eprint {http://arxiv.org/abs/2103.12439}
  {arXiv:2103.12439 [astro-ph.HE]} \BibitemShut {NoStop}%
\bibitem [{\citenamefont {Capozziello}\ \emph {et~al.}(2025)\citenamefont
  {Capozziello}, \citenamefont {Zare}, \citenamefont {Nieto},\ and\
  \citenamefont {Hassanabadi}}]{Capozziello:2023tbo}%
  \BibitemOpen
  \bibfield  {author} {\bibinfo {author} {\bibfnamefont {S.}~\bibnamefont
  {Capozziello}}, \bibinfo {author} {\bibfnamefont {S.}~\bibnamefont {Zare}},
  \bibinfo {author} {\bibfnamefont {L.~M.}\ \bibnamefont {Nieto}},  and
  \bibinfo {author} {\bibfnamefont {H.}~\bibnamefont {Hassanabadi}},\ }\href
  {\doibase 10.1016/j.dark.2025.102065} {\bibfield  {journal} {\bibinfo
  {journal} {\emph {Phys. Dark Univ.}}\ }\textbf {\bibinfo {volume} {50}},\
  \bibinfo {pages} {102065} (\bibinfo {year} {2025})},\ \Eprint
  {http://arxiv.org/abs/2311.12896} {arXiv:2311.12896 [gr-qc]} \BibitemShut
  {NoStop}%
\bibitem [{\citenamefont {Einstein}(1939)}]{Einstein:1939ms}%
  \BibitemOpen
  \bibfield  {author} {\bibinfo {author} {\bibfnamefont {A.}~\bibnamefont
  {Einstein}},\ }\href {\doibase 10.2307/1968902} {\bibfield  {journal}
  {\bibinfo  {journal} {\emph {Annals Math.}}\ }\textbf {\bibinfo {volume}
  {40}},\ \bibinfo {pages} {922} (\bibinfo {year} {1939})}\BibitemShut
  {NoStop}%
\bibitem [{\citenamefont {Geralico}\ \emph {et~al.}(2012)\citenamefont
  {Geralico}, \citenamefont {Pompi},\ and\ \citenamefont
  {Ruffini}}]{Geralico:2012jt}%
  \BibitemOpen
  \bibfield  {author} {\bibinfo {author} {\bibfnamefont {A.}~\bibnamefont
  {Geralico}}, \bibinfo {author} {\bibfnamefont {F.}~\bibnamefont {Pompi}},
  and \bibinfo {author} {\bibfnamefont {R.}~\bibnamefont {Ruffini}},\ }\href
  {\doibase 10.1142/S2010194512006356} {\bibfield  {journal} {\bibinfo
  {journal} {\emph {Int. J. Mod. Phys. Conf. Ser.}}\ }\textbf {\bibinfo
  {volume} {12}},\ \bibinfo {pages} {146} (\bibinfo {year} {2012})}\BibitemShut
  {NoStop}%
\bibitem [{\citenamefont {Konoplya}\ and\ \citenamefont
  {Zhidenko}(2022)}]{Konoplya:2022hbl}%
  \BibitemOpen
  \bibfield  {author} {\bibinfo {author} {\bibfnamefont {R.~A.}\ \bibnamefont
  {Konoplya}} and \bibinfo {author} {\bibfnamefont {A.}~\bibnamefont
  {Zhidenko}},\ }\href {\doibase 10.3847/1538-4357/ac76bc} {\bibfield
  {journal} {\bibinfo  {journal} {\emph {Astrophys. J.}}\ }\textbf {\bibinfo
  {volume} {933}},\ \bibinfo {pages} {166} (\bibinfo {year} {2022})},\ \Eprint
  {http://arxiv.org/abs/2202.02205} {arXiv:2202.02205 [gr-qc]} \BibitemShut
  {NoStop}%
\bibitem [{\citenamefont {Figueiredo}\ \emph {et~al.}(2023)\citenamefont
  {Figueiredo}, \citenamefont {Maselli},\ and\ \citenamefont
  {Cardoso}}]{Figueiredo:2023gas}%
  \BibitemOpen
  \bibfield  {author} {\bibinfo {author} {\bibfnamefont {E.}~\bibnamefont
  {Figueiredo}}, \bibinfo {author} {\bibfnamefont {A.}~\bibnamefont {Maselli}},
   and \bibinfo {author} {\bibfnamefont {V.}~\bibnamefont {Cardoso}},\ }\href
  {\doibase 10.1103/PhysRevD.107.104033} {\bibfield  {journal} {\bibinfo
  {journal} {\emph {Phys. Rev. D}}\ }\textbf {\bibinfo {volume} {107}},\
  \bibinfo {pages} {104033} (\bibinfo {year} {2023})},\ \Eprint
  {http://arxiv.org/abs/2303.08183} {arXiv:2303.08183 [gr-qc]} \BibitemShut
  {NoStop}%
\bibitem [{\citenamefont {Speeney}\ \emph {et~al.}(2024)\citenamefont
  {Speeney}, \citenamefont {Berti}, \citenamefont {Cardoso},\ and\
  \citenamefont {Maselli}}]{Speeney:2024mas}%
  \BibitemOpen
  \bibfield  {author} {\bibinfo {author} {\bibfnamefont {N.}~\bibnamefont
  {Speeney}}, \bibinfo {author} {\bibfnamefont {E.}~\bibnamefont {Berti}},
  \bibinfo {author} {\bibfnamefont {V.}~\bibnamefont {Cardoso}},  and \bibinfo
  {author} {\bibfnamefont {A.}~\bibnamefont {Maselli}},\ }\href {\doibase
  10.1103/PhysRevD.109.084068} {\bibfield  {journal} {\bibinfo  {journal}
  {\emph {Phys. Rev. D}}\ }\textbf {\bibinfo {volume} {109}},\ \bibinfo {pages}
  {084068} (\bibinfo {year} {2024})},\ \Eprint
  {http://arxiv.org/abs/2401.00932} {arXiv:2401.00932 [gr-qc]} \BibitemShut
  {NoStop}%
\bibitem [{\citenamefont {Pezzella}\ \emph {et~al.}(2025)\citenamefont
  {Pezzella}, \citenamefont {Destounis}, \citenamefont {Maselli},\ and\
  \citenamefont {Cardoso}}]{Pezzella:2024tkf}%
  \BibitemOpen
  \bibfield  {author} {\bibinfo {author} {\bibfnamefont {L.}~\bibnamefont
  {Pezzella}}, \bibinfo {author} {\bibfnamefont {K.}~\bibnamefont {Destounis}},
  \bibinfo {author} {\bibfnamefont {A.}~\bibnamefont {Maselli}},  and \bibinfo
  {author} {\bibfnamefont {V.}~\bibnamefont {Cardoso}},\ }\href {\doibase
  10.1103/PhysRevD.111.064026} {\bibfield  {journal} {\bibinfo  {journal}
  {\emph {Phys. Rev. D}}\ }\textbf {\bibinfo {volume} {111}},\ \bibinfo {pages}
  {064026} (\bibinfo {year} {2025})},\ \Eprint
  {http://arxiv.org/abs/2412.18651} {arXiv:2412.18651 [gr-qc]} \BibitemShut
  {NoStop}%
\bibitem [{\citenamefont {Bhowmik}\ \emph {et~al.}(2026)\citenamefont
  {Bhowmik}, \citenamefont {Chowdhury},\ and\ \citenamefont
  {Chakrabarti}}]{Bhowmik:2026owi}%
  \BibitemOpen
  \bibfield  {author} {\bibinfo {author} {\bibfnamefont {A.}~\bibnamefont
  {Bhowmik}}, \bibinfo {author} {\bibfnamefont {A.}~\bibnamefont {Chowdhury}},
  and \bibinfo {author} {\bibfnamefont {S.}~\bibnamefont {Chakrabarti}},\
  }\Eprint {http://arxiv.org/abs/2608.07678} {arXiv:2608.07678 [gr-qc]}
  \BibitemShut {NoStop}%
\bibitem [{\citenamefont {Cardoso}\ \emph {et~al.}(2009)\citenamefont
  {Cardoso}, \citenamefont {Miranda}, \citenamefont {Berti}, \citenamefont
  {Witek},\ and\ \citenamefont {Zanchin}}]{Cardoso:2008bp}%
  \BibitemOpen
  \bibfield  {author} {\bibinfo {author} {\bibfnamefont {V.}~\bibnamefont
  {Cardoso}}, \bibinfo {author} {\bibfnamefont {A.~S.}\ \bibnamefont
  {Miranda}}, \bibinfo {author} {\bibfnamefont {E.}~\bibnamefont {Berti}},
  \bibinfo {author} {\bibfnamefont {H.}~\bibnamefont {Witek}},  and \bibinfo
  {author} {\bibfnamefont {V.~T.}\ \bibnamefont {Zanchin}},\ }\href {\doibase
  10.1103/PhysRevD.79.064016} {\bibfield  {journal} {\bibinfo  {journal} {\emph
  {Phys. Rev. D}}\ }\textbf {\bibinfo {volume} {79}},\ \bibinfo {pages}
  {064016} (\bibinfo {year} {2009})},\ \Eprint {http://arxiv.org/abs/0812.1806}
  {arXiv:0812.1806 [hep-th]} \BibitemShut {NoStop}%
\bibitem [{\citenamefont {Jansen}(2017)}]{Jansen:2017oag}%
  \BibitemOpen
  \bibfield  {author} {\bibinfo {author} {\bibfnamefont {A.}~\bibnamefont
  {Jansen}},\ }\href {\doibase 10.1140/epjp/i2017-11825-9} {\bibfield
  {journal} {\bibinfo  {journal} {\emph {Eur. Phys. J. Plus}}\ }\textbf
  {\bibinfo {volume} {132}},\ \bibinfo {pages} {546} (\bibinfo {year}
  {2017})},\ \Eprint {http://arxiv.org/abs/1709.09178} {arXiv:1709.09178
  [gr-qc]} \BibitemShut {NoStop}%
\bibitem [{\citenamefont {Cardoso}\ \emph {et~al.}(2018)\citenamefont
  {Cardoso}, \citenamefont {Costa}, \citenamefont {Destounis}, \citenamefont
  {Hintz},\ and\ \citenamefont {Jansen}}]{Cardoso:2017soq}%
  \BibitemOpen
  \bibfield  {author} {\bibinfo {author} {\bibfnamefont {V.}~\bibnamefont
  {Cardoso}}, \bibinfo {author} {\bibfnamefont {J.~L.}\ \bibnamefont {Costa}},
  \bibinfo {author} {\bibfnamefont {K.}~\bibnamefont {Destounis}}, \bibinfo
  {author} {\bibfnamefont {P.}~\bibnamefont {Hintz}},  and \bibinfo {author}
  {\bibfnamefont {A.}~\bibnamefont {Jansen}},\ }\href {\doibase
  10.1103/PhysRevLett.120.031103} {\bibfield  {journal} {\bibinfo  {journal}
  {\emph {Phys. Rev. Lett.}}\ }\textbf {\bibinfo {volume} {120}},\ \bibinfo
  {pages} {031103} (\bibinfo {year} {2018})},\ \Eprint
  {http://arxiv.org/abs/1711.10502} {arXiv:1711.10502 [gr-qc]} \BibitemShut
  {NoStop}%
\bibitem [{\citenamefont {Maggiore}\ \emph {et~al.}(2020)\citenamefont
  {Maggiore} \emph {et~al.}}]{ET:2019dnz}%
  \BibitemOpen
  \bibfield  {author} {\bibinfo {author} {\bibfnamefont {M.}~\bibnamefont
  {Maggiore}} \emph {et~al.} (\bibinfo {collaboration} {ET}),\ }\href {\doibase
  10.1088/1475-7516/2020/03/050} {\bibfield  {journal} {\bibinfo  {journal}
  {\emph {JCAP}}\ }\textbf {\bibinfo {volume} {03}},\ \bibinfo {pages} {050}
  (\bibinfo {year} {2020})},\ \Eprint {http://arxiv.org/abs/1912.02622}
  {arXiv:1912.02622 [astro-ph.CO]} \BibitemShut {NoStop}%
\bibitem [{\citenamefont {Tulin}\ and\ \citenamefont
  {Yu}(2018)}]{Tulin:2017ara}%
  \BibitemOpen
  \bibfield  {author} {\bibinfo {author} {\bibfnamefont {S.}~\bibnamefont
  {Tulin}} and \bibinfo {author} {\bibfnamefont {H.-B.}\ \bibnamefont {Yu}},\
  }\href {\doibase 10.1016/j.physrep.2017.11.004} {\bibfield  {journal}
  {\bibinfo  {journal} {\emph {Phys. Rept.}}\ }\textbf {\bibinfo {volume}
  {730}},\ \bibinfo {pages} {1} (\bibinfo {year} {2018})},\ \Eprint
  {http://arxiv.org/abs/1705.02358} {arXiv:1705.02358 [hep-ph]} \BibitemShut
  {NoStop}%
\bibitem [{\citenamefont {Eberhardt}\ and\ \citenamefont
  {Ferreira}(2025)}]{Eberhardt:2025caq}%
  \BibitemOpen
  \bibfield  {author} {\bibinfo {author} {\bibfnamefont {A.}~\bibnamefont
  {Eberhardt}} and \bibinfo {author} {\bibfnamefont {E.~G.~M.}\ \bibnamefont
  {Ferreira}},\ }\Eprint {http://arxiv.org/abs/2507.00705} {arXiv:2507.00705
  [astro-ph.CO]} \BibitemShut {NoStop}%
\bibitem [{\citenamefont {Chen}\ and\ \citenamefont
  {Kotla{\v{r}}{\'\i}k}(2023)}]{Chen:2023akf}%
  \BibitemOpen
  \bibfield  {author} {\bibinfo {author} {\bibfnamefont {C.-Y.}\ \bibnamefont
  {Chen}} and \bibinfo {author} {\bibfnamefont {P.}~\bibnamefont
  {Kotla{\v{r}}{\'\i}k}},\ }\href {\doibase 10.1103/PhysRevD.108.064052}
  {\bibfield  {journal} {\bibinfo  {journal} {\emph {Phys. Rev. D}}\ }\textbf
  {\bibinfo {volume} {108}},\ \bibinfo {pages} {064052} (\bibinfo {year}
  {2023})},\ \Eprint {http://arxiv.org/abs/2307.07360} {arXiv:2307.07360
  [gr-qc]} \BibitemShut {NoStop}%
\bibitem [{\citenamefont {Hassanabadi}\ and\ \citenamefont
  {Zare}(2026)}]{Hassanabadi:2026mmi}%
  \BibitemOpen
  \bibfield  {author} {\bibinfo {author} {\bibfnamefont {H.}~\bibnamefont
  {Hassanabadi}} and \bibinfo {author} {\bibfnamefont {S.}~\bibnamefont
  {Zare}},\ }\Eprint {http://arxiv.org/abs/2609.23750} {arXiv:2609.23750
  [gr-qc]} \BibitemShut {NoStop}%
\bibitem [{\citenamefont {Ylla}\ \emph {et~al.}(2026)\citenamefont {Ylla},
  \citenamefont {Makino}, \citenamefont {Tanaka}, \citenamefont {Ishibashi},\
  and\ \citenamefont {Yoo}}]{Ylla:2026ffv}%
  \BibitemOpen
  \bibfield  {author} {\bibinfo {author} {\bibfnamefont {A.~U.~P.}\
  \bibnamefont {Ylla}}, \bibinfo {author} {\bibfnamefont {K.}~\bibnamefont
  {Makino}}, \bibinfo {author} {\bibfnamefont {A.}~\bibnamefont {Tanaka}},
  \bibinfo {author} {\bibfnamefont {A.}~\bibnamefont {Ishibashi}},  and
  \bibinfo {author} {\bibfnamefont {C.-M.}\ \bibnamefont {Yoo}},\ }\href
  {\doibase 10.1088/1475-7516/2026/09/046} {\bibfield  {journal} {\bibinfo
  {journal} {\emph {JCAP}}\ }\textbf {\bibinfo {volume} {09}},\ \bibinfo
  {pages} {046} (\bibinfo {year} {2026})},\ \Eprint
  {http://arxiv.org/abs/2603.15979} {arXiv:2603.15979 [gr-qc]} \BibitemShut
  {NoStop}%
\end{thebibliography}%

\end{document}